\documentclass[pdftex,12pt,twoside,footsepline,headsepline,a4paper,final]{scrbook}
\usepackage[verbose,a4paper,bottom=25mm,top=25mm]{geometry} 
\usepackage{scrpage2}
\usepackage[T1]{fontenc}
\usepackage[utf8x]{inputenc}
\usepackage{gensymb}
\usepackage{graphicx} 
\usepackage{multicol} 
\usepackage{floatrow}
\usepackage{amsmath}
\usepackage{amssymb}
\usepackage{booktabs}
\usepackage{braket}
\usepackage{pdfpages}

\usepackage{xcolor}
\usepackage{lipsum}
\definecolor{tcolor}{RGB}{189,96,5}
\usepackage{quotchap}
\colorlet{chaptergrey}{tcolor!50}

\newcommand*\diff{\mathrm{d}}

\newcommand*\ope[1]{\mathop{}\!\hat{#1}}
\newcommand*\nope[1]{\mathop{}\!\hat{#1}^\dagger}
\newcommand*\cre{\nope{a}}
\newcommand*\ann{\ope{a}}
\newcommand*\per{\mathrm{per}}
\newcommand*\e{\mathrm{e}}
\newcommand*\im{\mathrm{i}}

\usepackage{standalone}

\begin{document}

\frontmatter 

\newgeometry{left=0cm,right=0cm,top=2cm,bottom=2.6cm}
\begin{titlepage}


\center 
 
\includegraphics[width=0.7\textwidth]{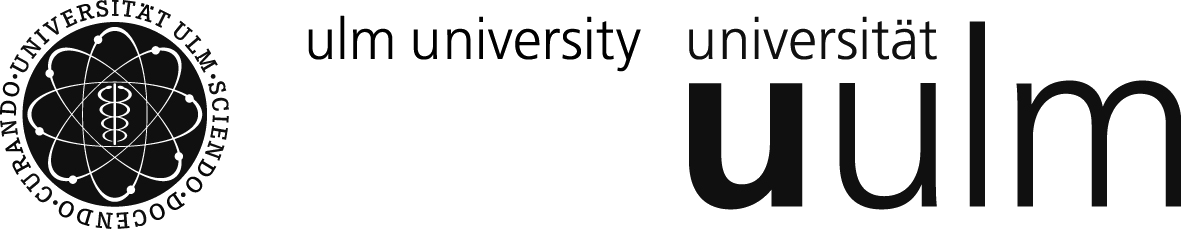}\\[1cm]
\begin{flushright}
\begin{minipage}[c]{4.5cm}
\large \textbf{Faculty of \\Natural Sciences}\\
Institute of \\Quantum Physics\hspace{2cm}\\[2cm]
\end{minipage}
~
\begin{minipage}{2.5cm}
\qquad
\end{minipage}
\end{flushright}


\begin{center}
           \noindent\colorbox{tcolor!50}{
           \begin{minipage}[c]{\paperwidth}
            {  \qquad \\
            \begin{center}
            \Huge \textsc{\bfseries Multi-Photon Interference\\ and Metrology\\}
            \end{center}
             \qquad\\}
             \end{minipage}
            }
\end{center} 
\textsc{\Huge Bachelor Thesis}\\
\qquad\\[4cm]

\begin{flushleft}\large
\hspace{2.5cm}Submitted by: Olaf Zimmermann \\
\hspace{2.5cm}olaf.zimmermann@uni-ulm.de\\
\hspace{2.5cm}Date: 12. June 2015\\[1cm]
\hspace{2.5cm}Examiner: Prof. Dr. Wolfgang P. Schleich\\
\hspace{2.5cm}Supervisor: Dr. Vincenzo Tamma
\end{flushleft}

\end{titlepage}

 \restoregeometry
\begin{quote}
\begin{center}
Eendoogs steeg de Aap vun'n Boom\\
(Dat g{\"u}ng em wull to good),\\
trock sik B{\"u}x un Stevel an\\
un leep sietdem to Foot.\\[0.5cm]

Nu weer he Mensch, b{\"o}{\"o}t F{\"u}{\"u}r an\\
un smurgel sik wat Kruut.\\
Denn strapaseer he sien Grips\\
un spikeleer wat ut.\\[0.5cm]

He mook sik {\"A}xt un Spier ut Steen.\\
He trock ut to'n Jogen\\
op Aanten, Haas un wilde Swien.\\
So kreeg he wat in'n Mogen.\\[0.5cm]

He bu sik Huus un Hoff un Borg.\\
He woor klook un kl{\"o}ker,\\
fl{\"o}{\"o}g as'n Vogel d{\"o}r de Luft.\\
Blot sien Welt kreeg L{\"o}cker.\\[0.5cm]

He hett PS, he hett Atom.\\
He dr{\"o}{\"o}mt (as ik un du).\\
Wenn de B{\"o}{\"o}m fehlt un de Bl{\"o}{\"o}m...\\
Eendoogs frogt he: Wat nu?\\[1.5cm]

Boy Lornsen: \textit{Eendoogs}. Sinfunikunzert.\\ Quickborn-Verlag, 2007.
\end{center}

\end{quote}\newpage
\section*{Abstract}
\begin{quote}
Recently, Motes et al. proposed in Phys. Rev. Lett. \textbf{114}, 170802 (2015) a linear optics interferometer with N identical single photon input states as a tool for sub-shot-noise phase estimation which does not require NOON states sources. 
This thesis investigates how the degree of distinguishability between the input photonic spectra affects the interferometric phase sensitivity, by employing the results obtained by Tamma and Laibacher in Phys. Rev. Lett. \textbf{114}, 243601 (2015). 
Remarkably, the obtained results show, that for small phases, the phase sensitivity in the case of completely distinguishable photons is identical to the one in the case of identical photons considered by Motes et al. apart from a constant factor that is independent of the number N of input photons. Consequently, multiphoton interference does not provide in the scheme proposed by Motes et al. any sub-shot-noise sensitivity for a large number N of photons.

\end{quote}

\let\cleardoublepage\clearpage
\tableofcontents
\let\cleardoublepage\clearpage
\mainmatter

\chapter{Introduction}

The applications of interference and more recently quantum interference in metrology measurements are numerous. However, approaches to super-sensitive measurements with precision close to the Heisenberg limit, the fundamental barrier for quantum optical metrology, have so far been reliant on the preparation of highly entangled states that are extremely hard and resource intensive to produce. 
In a recent paper \cite{Motes} Motes et al. proposed an alternative to these approaches by introducing the so called Quantum Fourier transform interferometer (QuFTI), claimed to perform phase measurements with a precision close to the Heisenberg limit while it only relies on the preparation of single-photon inputs, without the need of entangled sources.
The high fidelity supposedly arises naturally from the interference of indistinguishable photons in the interferometer that generates a high degree of entanglement. \\

In the presented thesis the QuFTI is examined in order to analyse, how the precision of the phase measurement is influenced by different degrees of distinguishability of the interferometer input. For this purpose a protocol recently introduced by Vincenzo Tamma and Simon Laibacher \cite{Tamma} is used. The procedure allows the description of multi-photon interference in arbitrary interferometers with an arbitrary number of single input photons with arbitrary degree of overlap in the photonic spectra.\\

In section \ref{sec: doc} a numerical approach is employed to determine the sensitivity of the QuFTI depending on the spectral overlap between the input photons.
The obtained results indicate that the sensitivity in the measurement of small phases with a  QuFTI is not influenced by the degree of distinguishability of the input single photons for a large number $n$ of photons. 
In the subsequent section \ref{proof} these results are confirmed analytically. 
In particular, it is shown that the phase sensitivities for completely distinguishable and completely indistinguishable photons differ only by a constant factor, that is independent of the number n of input photons $n$. 
Therefore the presented work proves that multi-photon indistinguishability and interference can not provide any sub-shot-noise sensitivity for large values of n in the QuFTI interferometer proposed in [1].
\chapter{Theoretical Foundations}
The present chapter discusses the physical and mathematical foundations necessary for the comprehension of the argumentation in this thesis. The basics of quantum optics and the description of light in the quantum regime are introduced starting from the electromagnetic field. Subsequently the quantum optical description of interference in passive linear interferometers and its possible applications in metrology are discussed.

\section{Quantization of the Electromagnetic Field}

This section briefly discusses the quantized representation of the electromagnetic field, starting from the classical model. The description is in principle based on the derivations given in \cite{Loudon,knight}.
In classical electrodynamics the electromagnetic field is most generally described by the vector potential $\vec{A}$ and the scalar potential $\Phi$, that generate the electric field $\vec{E}$ and the magnetic field $\vec{B}$ according to
\begin{align}
\vec{E}&=-\nabla\Phi -\frac{\partial}{\partial t}\vec{A}
\mathrm{\qquad and \qquad}
\vec{B}=\nabla\times\vec{A}.
\label{TQEFpotentiale}
\end{align} 
For the description of the radiation field the Coulomb gauge with the gauge condition $\nabla\cdot\vec{A}=0$ is applied and the absence of charges and currents is assumed. With this conditions $\Phi$ vanishes and the radiation is fully described by $\vec{A}$, which is expressed in the form of its modes. 
Each mode is defined by a wave vector $\vec{k}$ and the polarization direction $\vec{e}_{k\lambda}$, where the index $\lambda=1,2$ denotes one of the two possible polarizations orthogonal to the propagation direction. 
For simplicity the description will be limited to one direction, such that one mode is sufficiently described by the absolute value $k=|\vec{k}|$ of the wave vector or alternatively by the corresponding angular frequency $\omega_k=k c$, where $c$ denotes the speed of light in vacuum.
With these quantities the vector potential is defined by
\begin{equation}
\vec{A}=\vec{A}(\vec{r},t)=\sum\limits_k\sum\limits_{\lambda=1}^2 \vec{e}_{k\lambda} \left(A_{k\lambda}\e^{-\im\omega_{k}t}
\e^{i\vec{k}\vec{r}}+A^*_{k\lambda}\e^{\im\omega_{k}t}\e^{-i\vec{k}\vec{r}}\right)
\end{equation}
where the factor $A_{k\lambda}$ and its complex conjugate are the expansion coefficients of the given mode. 
The classical Hamiltonian $H$ of the radiation field follows from this description as \cite{Loudon}
\begin{equation}
H=\sum\limits_{k\lambda}
V\epsilon_0\omega^2_{k\lambda}\left( 
A_{k\lambda}A^*_{k\lambda}
+A^*_{k\lambda}A_{k\lambda}
\right) = \sum\limits_{k\lambda} H_{k\lambda}.
\end{equation}
The volume $V$ that appears in the Hamiltonian is the confining mode volume and $\epsilon_0$ is the permittivity of free space. At this point a mode is considered to be bound to a one dimensional cavity of length $L$. Therefore $V$ is the product of $L$ and the cross section $A$ in the plane orthogonal to $\vec{k}$.
This confinement imposes a standing wave condition on the modes in the cavity, allowing only $k$ with $k={2\pi n_{k}}/{L}$ and $ n_k \in \mathbb{N}$.
In order to quantize the Hamiltonian $H{k\lambda}$ of a single mode the canonical variables $q_{k\lambda}$ and $p_{k\lambda}$ are defined in terms of the coefficients $A_{k\lambda}$ and $A^*_{k\lambda}$ as
\begin{align}
q_{k\lambda}=\sqrt{V\epsilon_0 }(A_{k\lambda}+A^*_{k\lambda})
\mathrm{\qquad and \qquad}
\im p_{k\lambda}=\omega_{k} \sqrt{V\epsilon_0}(A_{k\lambda}-A^*_{k\lambda}).
\label{}
\end{align}

This definition brings $H_{k\lambda}$ into the form
\begin{equation}
H_{k\lambda}=\frac{1}{2}(p_{k\lambda}^2+\omega_{k}^2q_{k\lambda}^2)
\end{equation}
that is now quantized and associated with the respective Hamilton operator $\ope{H}_{k\lambda}$ by replacing $q_{k\lambda}$ and $p_{k\lambda}$ with the operators $\ope{q}_{k\lambda}$ and $\ope{p}_{k\lambda}$, leading to
\begin{equation}
\ope{H}_{k\lambda}=\frac{1}{2}(\ope{p}_{k\lambda}^2+\omega_{k}^2\ope{q}_{k\lambda}^2).
\label{TQEFHamOpe}
\end{equation}

$\ope{H}_{k\lambda}$ is equal to the Hamilton operator of an one dimensional quantum mechanical harmonic oscillator and it follows, that the eigenmodes of the quantized electromagnetic field can be described as harmonic oscillators with the same angular frequency. 
For the further description of these modes it is convenient to introduce the creation operator $\cre_{k\lambda}$ and the annihilation operator $\ann_{k\lambda}$ of the associated mode $k\lambda$ that are defined by
\begin{equation}
\ann_{k\lambda}=\sqrt{\frac{1}{2\hbar\omega_{k}}}(\omega_{k}\ope{q}_{k\lambda}+\im\ope{p}_{k\lambda})
\mathrm{\qquad and \qquad}
\cre_{k\lambda}=\sqrt{\frac{1}{2\hbar\omega_{k}}}(\omega_{k}\ope{q}_{k\lambda}-\im\ope{p}_{k\lambda}).
\end{equation}
$\cre_{k\lambda}$ and $\ann_{k\lambda}$ fulfil the commutation relation
\begin{equation}
[\ann_{k,\lambda},\cre_{k',\lambda'}]=\ann_{k,\lambda}\cre_{k',\lambda'}
-\cre_{k',\lambda'}\ann_{k,\lambda}=\delta_{k'k}\delta_{\lambda',\lambda}.
\label{TQEFcommrel}
\end{equation}
Introducing the number operator $\ope{n}_{k\lambda}=\cre_{k\lambda}\ann_{k\lambda}$, $\ope{H}_{k\lambda}$ becomes
\begin{equation}
\ope{H}_{k\lambda}=\frac{\hbar\omega_k}{2}(\ann_{k\lambda}\cre_{k\lambda}+\cre_{k\lambda}\ann_{k\lambda})=\hbar\omega_k(\ope{n}_{k\lambda}+\frac{1}{2}).
\label{TQEFHamOpediscrete}
\end{equation}

\section{The Photon}

After establishing the quantized description of the electromagnetic field, the photons are now introduced as the quanta of this field. 
These quanta are excitations of the eigenmodes or number states $\ket{n_{k\lambda}}$ of the field, that fulfil the eigenvalue relation
\begin{equation}
\ope{H}_{k\lambda}\ket{n_{k\lambda}}=\hbar\omega_k(\ope{n}_{k\lambda}+\frac{1}{2})\ket{n_{k\lambda}}=\hbar\omega_k(n_{k\lambda}+\frac{1}{2})\ket{n_{k\lambda}}.
\end{equation}
The defining eigenvalue $n_{k\lambda}$ is generally interpreted as the number of photons in the mode \cite[p. 133-138]{Loudon}.
In this representation the lowest possible state, the vacuum state $\ket{0}$, contains energy but no photons.
 This energy $E_0=\frac{1}{2}\hbar\omega_k$ is known as the zero point energy or vacuum energy. 
 When discussing interferometers, the zero point energy is the reason to consider inputs, that are in the vacuum state. 
Starting from the vacuum state, photons can be created or annihilated in a mode by the action of $\ann_{\lambda k}$ and $\cre_{\lambda k}$. The operators act according to 
\begin{equation}
\ann_{k\lambda}\ket{n_{k\lambda}}=\sqrt{n_{k\lambda}}\ket{n_{k\lambda}-1}
\mathrm{\qquad and \qquad}
\cre_{k\lambda}\ket{n_{k\lambda}}=\sqrt{n_{k\lambda}+1}\ket{n_{k\lambda}+1}
\end{equation} 
on arbitrary number states. For the vacuum state the annihilation operator acts according to $\ann_{k\lambda}\ket{0_{k\lambda}}=0$.
General states $\ket{\Psi_{k\lambda}}$ of single modes are expressed as linear superpositions of the number states
or powers of the creation operator
\begin{equation}
\ket{\Psi_{k\lambda}}=\sum\limits_{n_{k\lambda}} \alpha_{n_{k\lambda}} \ket{n_{k\lambda}}=\sum\limits_{n_{k\lambda}} \alpha_{n_{k\lambda}} \frac{(\cre_{k\lambda})^{n_{k\lambda}}}{\sqrt{n!}} \ket{0}.
\label{TPPsiSM}
\end{equation}

\subsubsection{Continuous Modes}

The previously mentioned confinement to a one dimensional cavity only allows a discrete mode spectrum. In free space this confinement doesn't exist and the resulting continuous spectrum can be derived by taking the limit $L \rightarrow \infty$.
In this case, the distance $\Delta\omega$ between two modes tends to 0 and the discrete mode spectrum is converted into a continuous spectrum of modes. The discrete sum over all modes further becomes an integral over the angular frequencies $\omega$. 
The discrete mode creation and annihilation operators are further replaced by the operators $\ann(\omega)$ and $\cre(\omega)$ of the continuous modes that follow from their discrete counterparts by the relations
\begin{equation}
\ann_{k\lambda} \rightarrow \Delta\omega\ann(\omega) \mathrm{\qquad and\qquad} \cre_{k\lambda} \rightarrow \Delta\omega\cre(\omega).
\end{equation}

These operators obey commutation relations that read equivalent to eq. \eqref{TQEFcommrel}
\begin{equation}
[\ann(\omega_1),\cre(\omega_2)]=\delta(\omega_1-\omega_2)
\mathrm{\qquad and \qquad}
[\ann(t_1),\cre(t_2)]=\delta(t_1-t_2).
\label{TPHOTONcommrel}
\end{equation}

A given frequency distribution $\xi(\omega)$ now defines the wave package creation operator $\cre_\xi$ that is given by 
\begin{equation}
\cre_\xi=\int\diff\omega\xi(\omega)\cre(\omega).
\label{TPHOTONcreatecontinuous}
\end{equation}
For most applications it is common to consider $\xi(\omega)$ to be a Gaussian distributions of the form
\begin{equation}
\xi(\omega)=\left(\frac{2\tau^2}{\pi}\right)^{\frac{1}{4}} \exp\left[ -\im(\omega_0-\omega)t_0- (\omega_0-\omega)^2\tau^2\right]
\end{equation}
with coherence time $\tau$, emission time $t_0$ and central frequency $\omega_0$ \cite[p. 242f]{Loudon}. Alternatively the temporal distribution $\chi$ can be used, which is the Fourier transform of $\xi$. If $\xi$ is Gaussian, $\chi$ is also Gaussian with
\begin{equation}
\chi(t)= \left(\frac{1}{2\pi\tau^2}\right)^{1/4} \exp\left[-\frac{(t-t_0)^2}{4\tau^2} -i\omega_0 t  \right].
\label{TPHOTONtempdist}
\end{equation}

\section{Multi Photon Interference}\label{sec:Theory  Tamma}

\begin{figure}[H]
 \begin{center}
 \includegraphics[width=0.6\textwidth]{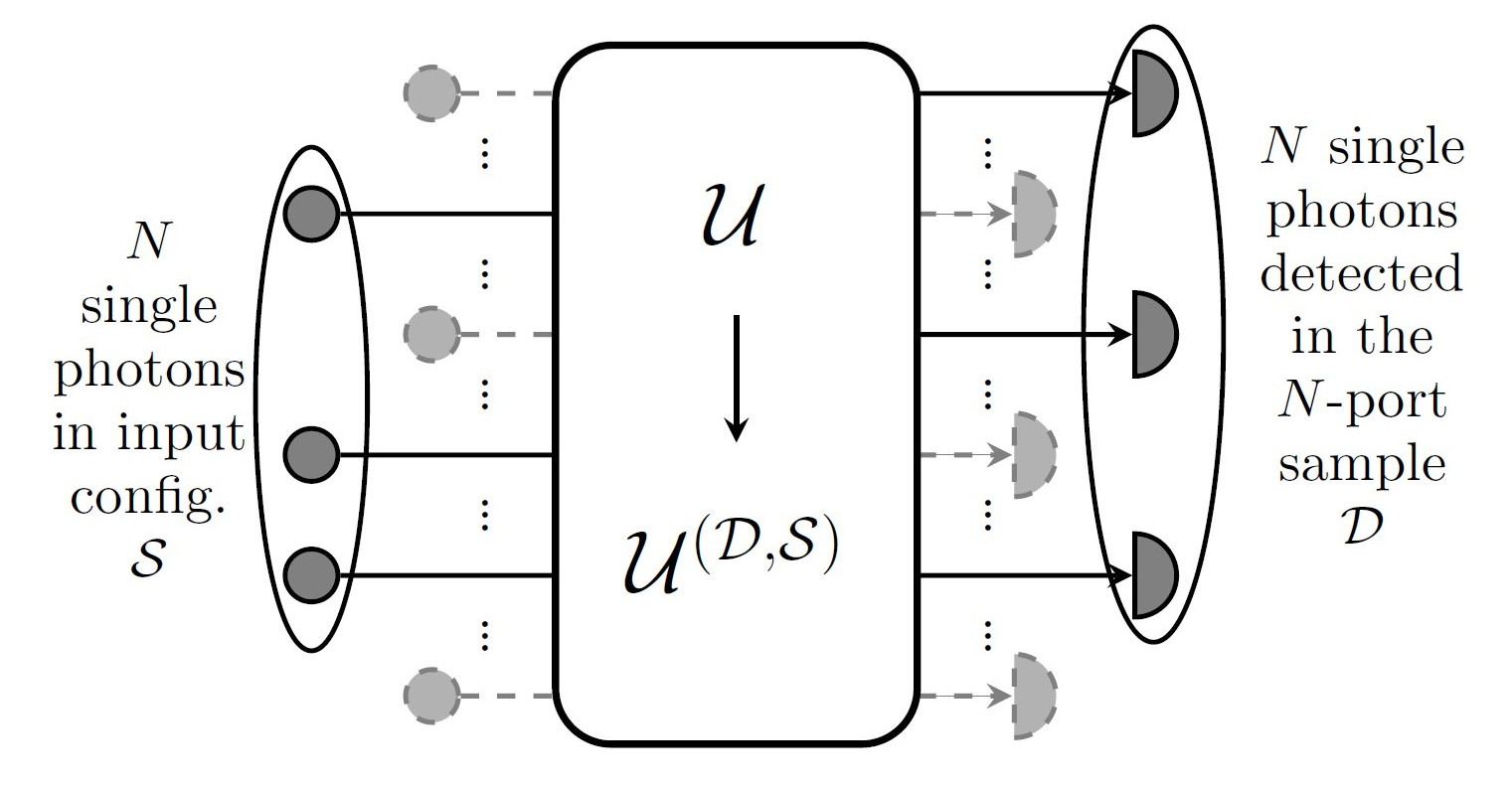}
\caption{A general passive linear interferometer, described by a unitary matrix $U$. The interferometer is fed with $N$ single input photons in a subset $S$ of all $M$ input ports. They can be detected in any possible sample $D$ of the $M$ output ports. For each output sample $D$ and input configuration $S$, the evolution in the interferometer is fully described by the submatrix $U^{(D,S)}$ \cite{Tamma}.}
\label{fig:setupTamma}
\end{center}
\end{figure}

In \cite{Tamma} Vincenzo Tamma and Simon Laibacher develop a general approach to describe multi-photon interference in arbitrary passive linear interferometers with single-photon input. 
In the following the results of this paper are outlined in order to apply them in the course of the work to the Quantum Fourier transform interferometer, which will be introduced shortly.\\

A passive linear interferometer with $M$ inputs and $M$ outputs, that is represented by figure \ref{fig:setupTamma}, is described by a $M\times M$ matrix $U$.
In the following single photon inputs in a subset $S$ of order $N<M$ are considered, each of which is described by a corresponding frequency distributions $\xi_s(\omega)$.
The polarizations are not taken into account, as all inputs are assumed to be polarized in the same direction. 
The input state of the interferometer is now the tensor product over all inputs $s$. 
\begin{equation}
\ket{S}= \bigotimes\limits_{s\in S} \ket{1[\xi_s]}_s \bigotimes\limits_{s\notin S} \ket{0}_s,
\label{Tinput}
\end{equation}
where the states $\ket{1[\xi_s]}$ are
\begin{equation}
\ket{1[\xi_s]}=\sum\limits_{\lambda=1,2}\int\limits_0^\infty \diff \omega \xi_s(\omega)\cre_{s,\lambda}(\omega)\ket{0}_s
\label{TMPIchiope}
\end{equation} 
in accordance with eq. \eqref{TPHOTONcreatecontinuous}.
The output after the evolution in the interferometer is a subset $D$ containing $N$ of all $M$ outputs. 
Now only the quantum paths connecting the entrance sample and the exit sample must be considered and it is sufficient to discuss the corresponding $N\times N$ submatrix $U^{(D,S)}$. 
Assuming narrow bandwidths of the entering photons and equal propagation times for all interferometer path, the authors determine the $N$-photon probability rate $G_{\{t_d,\vec{p}\}}^{(D,S)}=|\per T_{\{t_d\}}^{(D,S)}|^2$ with the matrices 
$T_{\{t_d\}}^{(D,S)}:=\left[ U_{d,s}\chi_s(t_d)\right]$,
that generally depend on the detection times $\{t_d\},d\in D$ of the interferometer output and the temporal distributions $\chi_s(t)$.\\

Quantum paths connecting the inputs and outputs are further shown to be indistinguishable for equal detection times and equal polarizations, although the entering photons are not necessarily identical. 
Therefore even the paths of distinguishable photons can interfere and path-distinguishability and photon-distinguishability must be distinguished. 
For \textit{non-resolving} measurements, that do not resolve the detection times or the polarization, the authors derive the averaged probability $P_{av}(D;S)$ to detect $N$ photons entering input ports in $S$ in the output ports of $D$ as
\begin{equation}
P_{av}(D;S)=
\sum\limits_{\rho\in \Omega_N} f_\rho(S) \per A_\rho^{(D,S)},
\label{TMPIProbAverage}
\end{equation}
 using the definition of a permanent eq. \eqref{Tpermanent} and defining the overlap factors 
\begin{equation}
f_{\rho}(S):=\prod\limits_{s \in S} \int\limits_{-\infty}^{+\infty} \diff t \chi_s^*(t)\chi_{\rho(s)}(t)
\label{TMPIIndFac}
\end{equation}
and the interference-type matrices
\begin{equation}
A_\rho^{(D,S)}=U^*_{d,s}U_{d,\rho(s)}.
\label{TAmatrix}
\end{equation} 

The probability is called  \textit{averaged} as the non-resolved parameters are averaged over the measurement interval. 
Two limiting scenarios for different degrees of distinguishability are considered. 
First, if all quantum paths are distinguishable all $f_{\rho}$  but the one corresponding to $\rho=\mathbf{1}$ vanish. The only contributing matrix is then $T^{(D,S)}=A_\mathbf{1}^{(D,S)}$ and the averaged probability $P_{T}(D;S)$ is accordingly 
\begin{equation}
P^{(T)}(D;S)=\per T^{(D,S)}.
\label{TprobNoOverlap}
\end{equation}
 For full path-indistinguishability, all $f_{\rho(S)}$ are equal to 1 and the probability $P_{U}(D;S)$ results in 
\begin{equation}
P^{(U)}(D;S)=\sum\limits_{\rho\in \Omega_N}\sum\limits_{\delta\in \Omega_N} \prod\limits_{s\in S} U^*_{\delta(s),s}U_{\delta(s),\rho(s)}
=|\per U^{(D,S)}|^2.
\label{TprobFullOverlap}
\end{equation}

\section{Entanglement}

According to \cite{Review} entanglement is best described as a property of more two or more quantum systems such that the state of one system cannot be regarded without considering the other systems. A simple example, that also illustrates the schemes of the previous section, is given by two single photons entering a lossless 50:50 beam splitter from two adjacent sides \cite{ShivA,HOM}. The beam splitter is described by
\begin{equation}
U_{BS}=\frac{1}{\sqrt{2}}\begin{pmatrix}
1 & \im \\
-\im & -1
\end{pmatrix}
\end{equation}
\cite[p. 89-91]{Loudon}.
With the tools introduced before, the probability $P_{av}$ to detect one photon per mode is
\begin{align}
P_{av}=&\per T+ f_{\{2,1\}} \per A_{\{2,1\}} =\frac{1}{2}-\frac{1}{2}
\left(\int\limits_{-\infty}^{+\infty} \diff t \chi_1^*(t)\chi_{2}(t)\right)^2
\end{align}
according to eq. \eqref{TMPIProbAverage}.
Here $\{2,1\}$ describes the permutation that interchanges the two indices. If we consider for example Gaussian distributions as given by \eqref{TPHOTONtempdist} with emission times $t_{01}$ and $t_{02}$, $f_{\{2,1\}}$ becomes
\begin{align}
 f_{\{2,1\}}&=\exp\left(-\dfrac{\Delta t_{1,2}^2}{4\tau^2}\right)
\end{align}
with $\Delta t_{1,2}=t_{01}-t_{02}$. $P_{av}$ is then
\begin{align}
P_{av}=\frac{1}{2}-\frac{1}{2}\exp\left(-\dfrac{\Delta t_{1,2}^2}{4\tau^2}\right).
\end{align}

\begin{figure}[!htb]
 \begin{center}
 \includegraphics[width=0.5\textwidth]{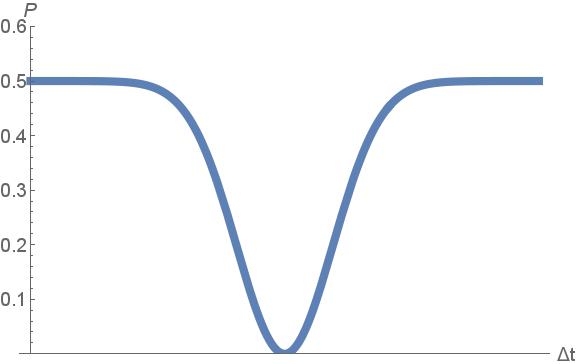}
\caption{ Probability distribution for a two-fold detection event in a lossless 50:50 beam splitter, assuming single photon inputs with Gaussian temporal distributions with variable emission times. The probability is plotted over emission time difference $\Delta t_{1,2}=t_{01}-t_{02}$. }
\label{HOMplot}
\end{center}
\end{figure}

Figure \eqref{HOMplot} shows $P_{av}$ as a function of $\Delta t_{1,2}$.
For equal emission times the probability vanishes
due to the quantum interference of the two possible paths and the output state $\ket{D}$ is accordingly a superposition 
\begin{equation}
\ket{D}=\frac{\ket{20}+\ket{02}}{\sqrt{2}}
\end{equation}
of the two modes with 2 photons in one of the output ports.\\

If the system is in this state and one of the photons is measured in one of the output ports, the other photon exit the same port. Therefore the state is entangled. 
If the paths are fully distinguishable, the probability for a 2-fold detection is $1/2$ as seen in the limit $\Delta t_{1,2} \rightarrow \pm\infty$ and the detection of one photon allows no conclusion about the other photon. 
This basic example shows that entanglement can be generated by quantum optical interference and consequently the magnitude of distinguishability is directly linked to the magnitude of generated entanglement.

\section{Application of Interference in Metrology }\label{sec:Theory Motes}

Quantum interference is a fundamental tool in metrology. The applications of the resulting measurement techniques in science are multiple, the basic principle however is simple: A measurement that is based on interference comes down to the determination of a phase shift $\varphi$ between different beams of light, that is caused by some feature of the quantity to be measured. The quantity is then determined from $\varphi$. Because of this relation, the precision of the phase estimation is directly linked to the precision of the desired quantity.\\

The fundamental restrictions on the precision of metrology schemes follow from the Heisenberg uncertainty relation\cite{Ou}.
For interference measurements the uncertainty relation connects the phase sensitivity $\Delta\varphi$ with the fluctuations $\Delta N$ of the photon number $N$ as $\Delta \varphi\Delta N\ge 1$. 
The uncertainty $\Delta N$ is therefore a lower bound for $\Delta \varphi$ according to $\Delta \varphi\ge {1}/{\Delta N}$.
For classical measurements the best feasible fluctuation to photon number ratio is $(\Delta N)^2 \propto \braket{N}$, such that the lower bound 
\begin{equation}
\Delta \varphi_{SNL}\ge \frac{1}{\sqrt{\braket{N}}}.
\end{equation}
is the shot noise limit $\Delta \varphi_{SNL}$ \cite{Dowling}.
For measurements that employ entanglement and quantum interference, the final restriction to the precision is set by the Heisenberg limit $\Delta \varphi_{HL}$ corresponding to $(\Delta N)^2 \propto \braket{N}^2$. 
The boundary then becomes
\begin{equation}
\Delta \varphi_{HL}\ge \frac{1}{\braket{N}}.
\label{Heisenberg}
\end{equation}
This is the reason to consider entanglement the most valuable resource for quantum optical metrology.\\

\section{The Quantum Fourier Transform Interferometer}
\begin{figure}[!htb]
 \begin{center}
 \includegraphics[width=0.6\textwidth]{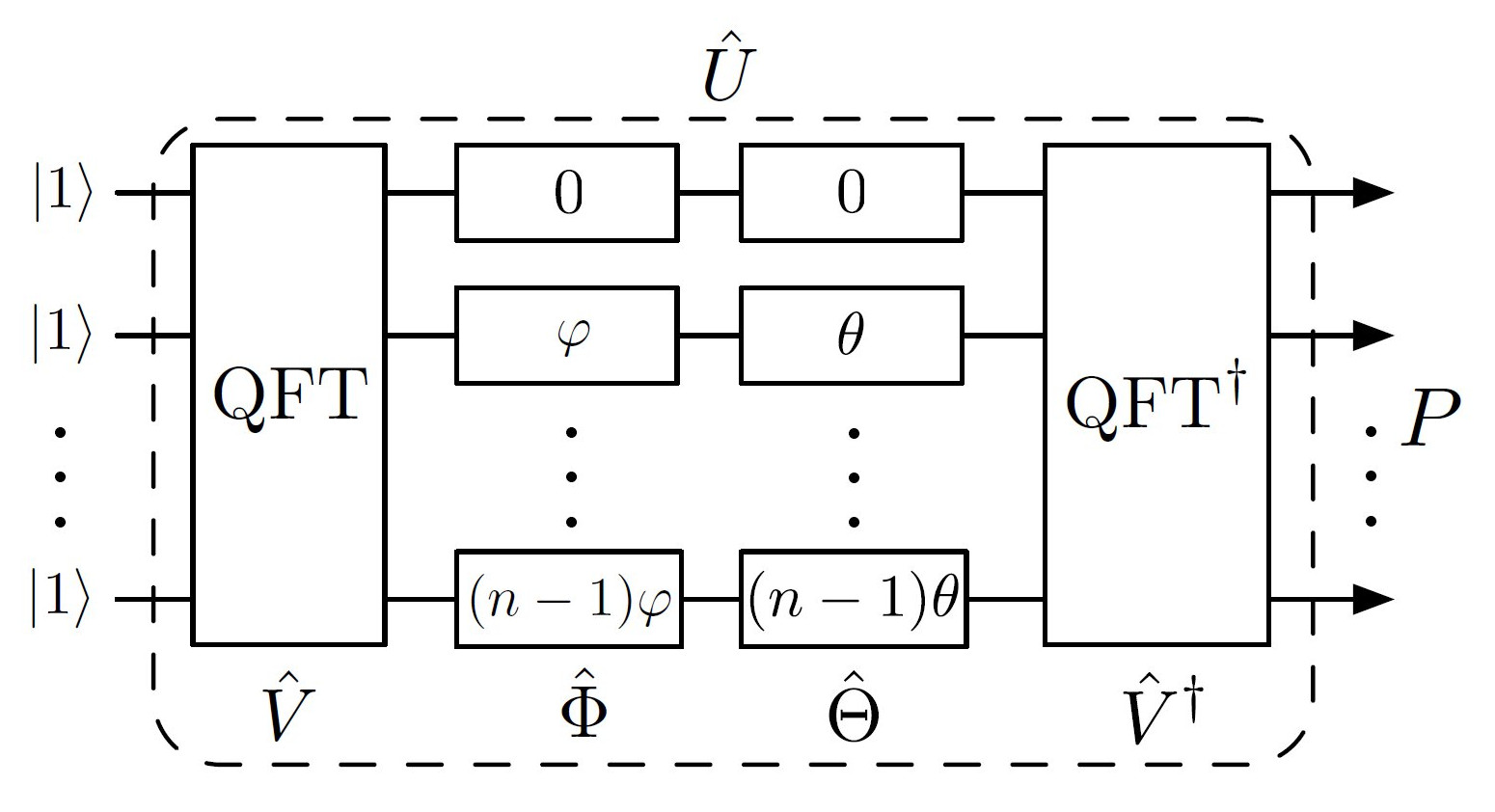}
\caption{Graphical representation of the QuFTI. The full interferometer is described by the unitary matrix $U=V\Phi\Theta V^\dagger$ where $\Phi$ is an unknown linear phase gradient to be measured. $\Theta$ is a second phase gradient used for the calibration of the device and $V$ and $V^\dagger$ are the quantum Fourier transformation and it's hermitian conjugate \cite{Motes}.}
\label{fig:QuFTI}
\end{center}
\end{figure}

An approach to sub-shotnoise measurements using only single photon inputs, was recently proposed in \cite{Motes}. 
The introduced quantum Fourier transform interferometer (QuFTI) is described by a matrix $U^{(n)}$, that is based on a phase shifter $\Phi$ represented by a diagonal matrix in phase space. 
The elements of $\Phi$ are linearly increasing phase shifts along the diagonal according to
\begin{equation}
\Phi_{j,k}=\delta_{j,k}\exp[\im(j-1)\varphi].
\label{TPhaseShifter}
\end{equation}
Additionally another phase shifter $\Theta$ is defined similarly by the elements
\begin{equation}
\Theta_{j,k}=\delta_{j,k}\exp[\im(j-1)\theta],
\end{equation}
that is used for the calibration of the device by shifting the phase angle range to the optimal measurement regime. 
The following discussion assumes the phase angle $\varphi$ to be already in the optimal regime. Therefore $\theta$ is set to zero and $\Theta$ is the identity matrix.
In order to describe the QuFTI in state space, the $n$-mode quantum Fourier transformation is applied to $\Phi$. 
This transformation is defined by the matrix $V$ with the elements
\begin{equation}
V_{j,k}^{(n)}=\frac{1}{\sqrt{n}}\exp\left(\frac{-2\im jk\pi}{n}\right)
\end{equation}
and its hermitian conjugate. 
The QuFTI matrix $U^{(n)}$ is then the matrix product $U^{(n)}=V\Phi\Theta V^\dagger=V\Phi V^\dagger$ with the elements
\begin{align}
U_{j,k}^{(n)}&=
\sum\limits_{l=1}^n V_{j,l}\Phi_{l,l}V_{l,k}^\dagger=
\sum\limits_{l=1}^n \frac{1}{n} \exp\left(\frac{-2\im jl\pi}{n}\right) \exp[\im(l-1)\varphi] \exp\left(\frac{2\im lk\pi}{n}\right)
%
\nonumber\\&\overset{\eqref{TgeoSeries}}{=}
\frac{1-e^{\im n\varphi}}{n\left(e^{2\im\pi(j-k)/n}-e^{\im\varphi}\right)}.
\label{TMatrixElementsU}
\end{align}

The authors investigated the probability for measuring one photon per output in the QuFTI given an input of $n$ fully indistinguishable photons, each  entering a different port. 
In accordance with eq. \eqref{TprobFullOverlap}, this probability is  $P=|\per U^{(n)}|^2$ and the uncertainty of the phase measurement follows from the uncertainty $\Delta P=\sqrt{P-P^2}$ of the probability as
\begin{equation}
\Delta\varphi=\frac{\sqrt{P-P^2}}{|\frac{\partial P}{\partial \varphi}|}.
\label{Tphaseuncertainty}
\end{equation}
using the Gaussian error propagation formula \cite[p. 858]{Bronstein}.\\

The authors were able to detect a numerical pattern 
\begin{equation}
\per U^{(n)}=\frac{1}{n^{n-1}}\prod\limits_{j=1}^{n-1} \left[j\e^{\im n\varphi}+n-j\right]
\label{TpatternU}
\end{equation}
 when calculating $\per U^{(n)}$ for $n$ up to 25. 
$P_n^{(U)}$ was then derived as
\begin{equation}
P_n^{(U)}=\left|\per U^{(n)}\right|^2=\frac{1}{n^{2n-2}} \prod\limits_{j=1}^{n-1} \left[a_n(j)\cos(n\varphi)+b_n(j)\right]
\label{TpatternProb}
\end{equation}
with the factors $a_n(j)=2j(n-j)$ and $b_n(j)=n^2-2nj+2j^2$.
In order to get a measure for the magnitude of the phase sensitivity the approximation $\cos(\varphi)\approx 1-1/{2}\varphi^2$ was applied to eq. \eqref{TpatternProb} to calculate $\Delta\varphi$ for small phases $\varphi$.
This yields the probability 
\begin{align}
 P=&1-\varphi^2\frac{n(n-1)(n+1)}{6}=1-k(n)\varphi^2
\label{TProbSAA}
\end{align}
as derived in the appendix \ref{appendixPUn} and $\Delta\varphi$ is  
\begin{align}
\Delta\varphi=\dfrac{\sqrt{1-k(n)\varphi^2-1+2k(n)\varphi^2}}{2k(n)|\varphi|}
= \dfrac{1}{2\sqrt{k(n)}}=
\sqrt{\frac{3}{n(n-1)(n+1)}}.
\label{TphaseSensitivitySAA}
\end{align}

\begin{figure}[!htb]
 \begin{center}
 \includegraphics[width=0.9\textwidth]{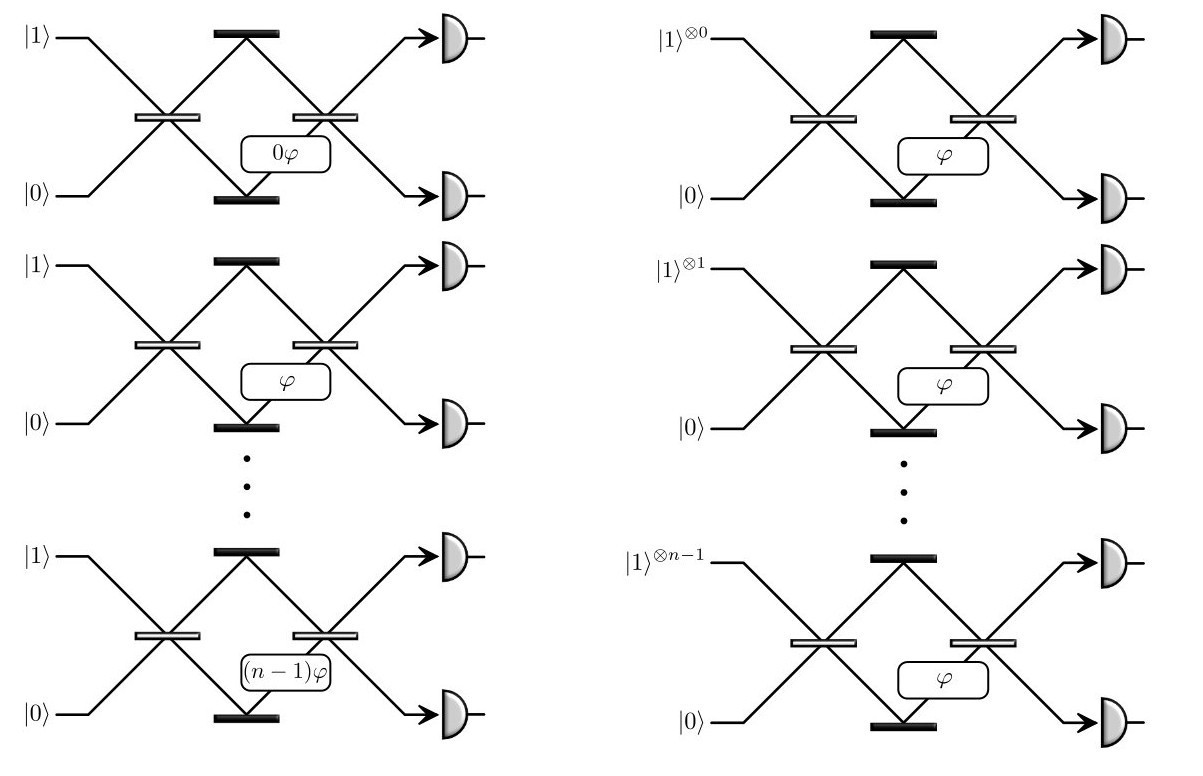}
\caption{Mach-Zehnder interferometer cascade (left) introduced in \cite{Motes} with linearly increasing phase shifts, each fed with single photon inputs. As one interrogation of a phase shift $k\varphi$ is equal to $k$ interrogations of the same phase shift $\varphi$, this setup is argued to be equivalent to a cascade of Mach-Zehnder interferometer (right) with an increasing number of input photons.}
\label{fig:MachZehnder}
\end{center}
\end{figure}

In order to construct a measure for the shot noise limit and the Heisenberg limit, the authors derived these limits from a supposedly comparable classical setup, consisting of a cascade of $n$ two mode Mach-Zehnder interferometers with the same linearly increasing phase shift sequence that is given by the phase shifter $\Phi$ in eq. \eqref{TPhaseShifter}.   
Each Mach-Zehnder interferometer is then fed with a single photon input into one entrance while the other entrance is in the vacuum state, thus keeping the input photons from interfering.
The authors further argue, that one interrogation in an interferometer producing a phase shift of $k\varphi$ is equivalent to $k$ interrogations in one interferometer with a phase shift $\varphi$, which is argued to be equivalent to $k$ uncorrelated photons entering the same port of a single interferometer. A graphical representation of the argumentation is given in figure \ref{fig:MachZehnder}.
 The number of interrogations $N_i$ in the setup which is given by 
\begin{equation}
N_i=\sum\limits_{k=0}^{n-1} k\overset{\eqref{TgaussianSum}}{=}\frac{n(n-1)}{2}
\end{equation}
is transferred into an equivalent photon number $N=N_i+1$, where the additional photon is the input photon that does not interrogate a phase shift. 
The shot noise limit is then defined by 
\begin{equation}
\Delta\varphi_{SNL}=\sqrt{\frac{1}{\frac{n(n-1)}{2} +1}}=\sqrt{\frac{2}{n(n-1)+2}}
\end{equation}
and the Heisenberg limit is defined as
\begin{equation}
\Delta\varphi_{HL}=\frac{2}{n(n-1)+2}.
\end{equation}
Therefore the authors argue that the precision of the QuFTI given in eq. \eqref{TphaseSensitivitySAA} beats the shotnoise limit.

\section{Mathematical Formulae}
\subsubsection{Permanent \cite[p. 1]{Permanent}}

The permanent of a $N\times N$ Matrix $A$ is defined as
\begin{equation}
\per A=\sum\limits_{\sigma\in \Omega_N} \prod\limits_{l=1}^n a_{l,\sigma(l)}
\label{Tpermanent}
\end{equation}

in which $\Omega_N$ is the symmetric group of order $N$ that contains all possible permutations $\sigma$ of the set $\{1,2,...,n\}$.

\subsubsection{Series Identities \cite[p. 19]{Bronstein}}

\begin{equation}
a_0 \sum\limits_{k=0}^{n} q^k=a_0\frac{1-q^{n+1}}{1-q}
\label{TgeoSeries}
\end{equation}

\begin{equation}
a_0 \sum\limits_{k=0}^{n} k\cdot q^k=a_0\frac{n\cdot q^{n+2}-(n+1)q^{n+1}+q}{(q-1)^2}
\label{TgeoSeries(2)}
\end{equation}

\begin{equation}
a_0 \sum\limits_{k=0}^{n} k=a_0\frac{n(n+1)}{2}
\label{TgaussianSum}
\end{equation}

\begin{equation}
a_0 \sum\limits_{k=0}^{n} k^2=a_0\frac{n(n+1)(2n+1)}{6}
\label{TgaussianSum(2)}
\end{equation}

\subsubsection{Trigonometric Series}

\begin{equation}
\sum\limits_{l=1}^{n-1}\frac{1}{\sin^2(\frac{\pi l}{n})}=\frac{n^2-1}{3} 
\label{TSinusId}
\end{equation}

The proof for the Identity has been found by Simon Laibacher in \cite{StackEx}. It is attached to this thesis in the appendix.
\chapter{Numerical Findings on the Influence of Distinguishability}\label{sec: doc}

In the present section the influence of the degree of distinguishability on the probability of an n-fold detection and the fidelity of the measurements is investigated numerically. 
For this purpose, the tools presented in section \ref{sec:Theory Tamma} are applied to the QuFTI matrix $U^{(n)}$ defined in eq. \eqref{TMatrixElementsU} using the numerical \textit{Mathematica} scripts in Appendix \ref{APPENDIXMathematica}.  

\section{Full Distinguishability \label{sec: FPD}}

The probability $P^{(T)}_n=\per T^{(n)}$ for an n-fold detection with fully indistinguishability input photons is according to to eq. \eqref{TprobNoOverlap} given by the permanent of the matrix $T^{(n)}$ with the elements 
\begin{equation}
T^{(n)}_{j,k}
= \left| U^{(n)}_{j,k} \right|^2= \left| \frac{1-e^{\im n\varphi}}{n\left(e^{2\im\pi(j-k)/n}-e^{\im\varphi}\right)} \right|^2
=\frac{1}{n^2} \frac{1-\cos(n\varphi)}
{1-\cos(2\pi\frac{j-k}{n}-\varphi)},
\end{equation}
that are the squared absolute values of the elements of $U^{(n)}$ in eq. \eqref{TMatrixElementsU} .\\

This representation of the matrix is implemented in a \textit{Mathematica} script, provided in the appendix \ref{APPENDIXMathematica}, to compute  $P^{(T)}_n$ for various numbers $n$ of input photons.
The script uses the standard implementation of the permanent calculation provided by \textit{Mathematica} \cite{MathematicaPer} and is based on a script by Alexander M{\"u}ller\footnote{Institute of Quantum Physics, University of Ulm.}. 
The probabilities $P^{(T)}_n$ are computed for $n=2,3,4,6,8$ as functions of the phase $\varphi$. 
The code fails to generate results for the permanents of larger matrixes as well as for odd numbers apart from $n=3$ due to the high computing time for simplifications of the permanents.
The probabilities in table \ref{tableNOpower} are expressed as polynomials of $\cos(n\varphi)$ in the form
\begin{equation}
P^{(T)}_n(\varphi)=\frac{1}{n^{2n-2}}\sum\limits_{j=0}^{n-1} a_j(n) \cos^j(n\varphi).
\label{docPTpower}
\end{equation}
In this representation the coefficients $a_j(n)$ fulfil
$
P^{(T)}_n(0)=1/{n^{2n-2}}\sum a_j(n)=1
$ 
but no further dependence of $a_j(n)$ on $n$ and $j$ is found. 
A pattern in the form of a product with whole numbers like the one observed for the fully indistinguishable case in eq. \eqref{TpatternU} is at least be ruled out by a prime factorization of the coefficients. \\

\renewcommand\arraystretch{1.5} 


\begin{table}
  \begin{center}
\begin{tabular}{|c||l|}
\hline
 n & $P^{(T)}_n $ \\\hline\hline
 2 & $\frac{1}{4} \left( 3+\text{Cos}[2 \varphi ]\right)$ \\\hline
 3 & $\frac{1}{3} \cdot \frac{1}{81} \left(143+92 \text{Cos}[3 \varphi ]+8 \text{Cos}[3 \varphi ]^2\right)$ \\\hline
 4 & $\frac{4}{4096}\left(481+449 \text{Cos}[4 \varphi ]+91 \text{Cos}[4 \varphi ]^2+3 \text{Cos}[4 \varphi ]^3\right)$ \\\hline
 6 & $\frac{8}{60466176}\left(2306819+3427228 \text{Cos}[6 \varphi ]+1563379 \text{Cos}[6 \varphi ]^2\right. $\\&   \qquad\qquad\qquad$\left.+249120 \text{Cos}[6 \varphi ]^3+11666 \text{Cos}[6 \varphi ]^4+60
\text{Cos}[6 \varphi ]^5\right)$ \\\hline
 8 & $\frac{256}{4398046511104}\left(3442375497+6960069125 \text{Cos}[8 \varphi ]+4984743781 \text{Cos}[8 \varphi ]^2\right. $\\&   \qquad\qquad\qquad$\left. +1563917761 \text{Cos}[8 \varphi ]^3+216734635 \text{Cos}[8
\varphi ]^4\right. $\\&   \qquad\qquad\qquad$\left.+11840831 \text{Cos}[8 \varphi ]^5+187239 \text{Cos}[8 \varphi ]^6+315 \text{Cos}[8 \varphi ]^7\right)$ \\ \hline
\end{tabular}
\end{center}
  \caption{Probabilities $P^{(T)}_n $ for fully distinguishable input photons, determined by the \textit{Mathematica} code in Appendix \ref{APPENDIXMathematica}.\label{tableNOpower}}
\end{table}

\begin{figure}[]
 \begin{center}
 \includegraphics[width=0.9\textwidth]{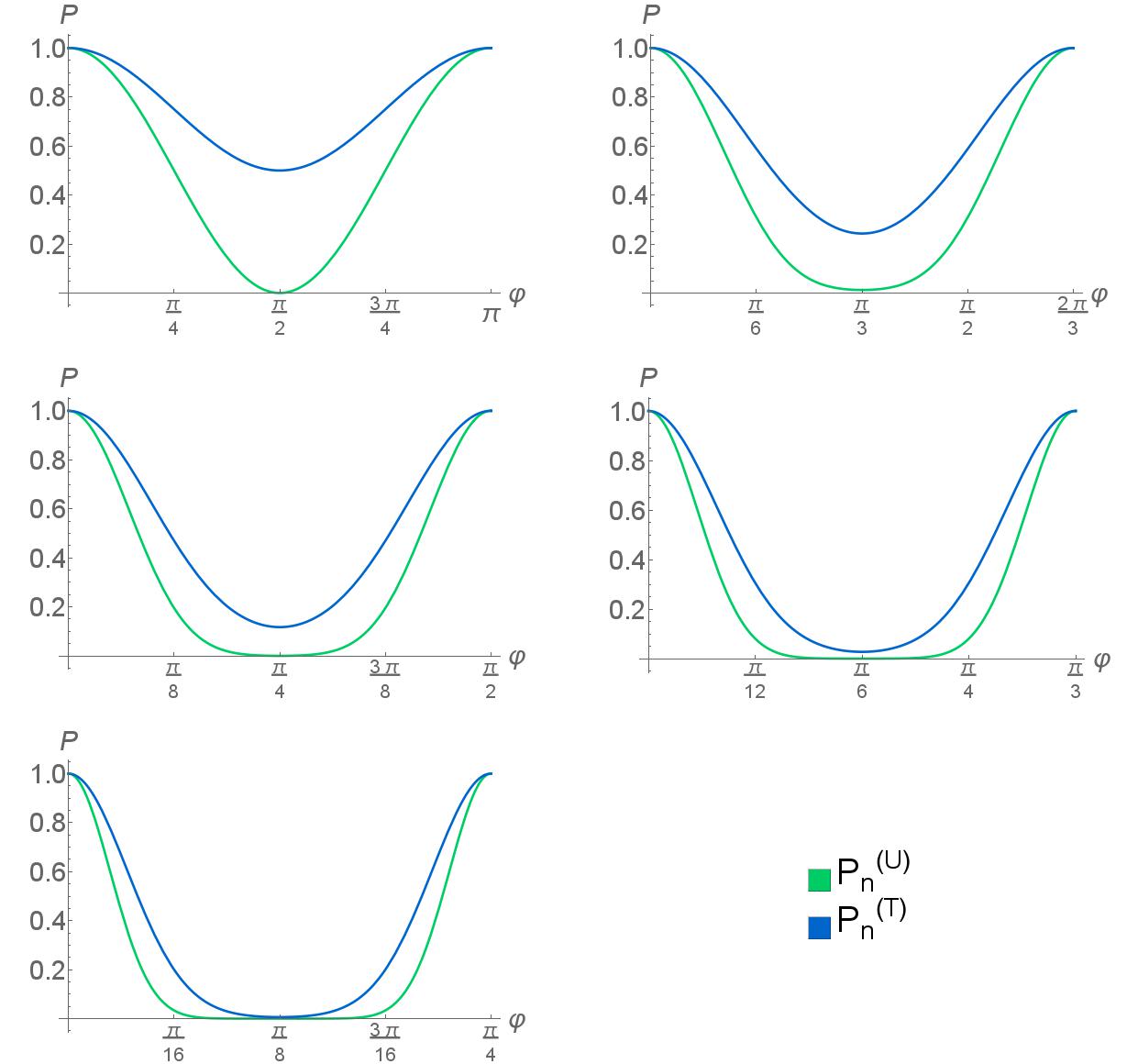}
\caption{Comparison between the probabilities to detect one photon per output of the QuFTI given fully distinguishable (blue) or fully indistinguishable (green) input photons. Both are plotted over the phase $\varphi$ for $n$=2,3,4,6,8. \label{figure:Probabilities} }
\end{center}
\end{figure}

\begin{table}[!htb]
  \begin{center}
\begin{tabular}{|c|c|c|c|c|c|}
\hline
 n & 2&3&4&6&8\\\hline
$P^{(T)}_n(\frac{\pi}{2n})$ &$\frac{1}{2} $&$ \frac{59}{243}$ &$ \frac{15}{128}$ &$\frac{12841}{472392}$& $\frac{845415}{134217728}$\\\hline
\end{tabular}
\end{center}
  \caption{Minima of $P^{(T)}_n$.\label{tableMinRelFr}}
\end{table}



Fig. \ref{figure:Probabilities} compares the probabilities $P^{(U)}_n$ and $P^{(T)}_n$ as a function of the phase $\varphi$. 
The plot shows a dip similar to the 2-photon interference in figure \ref{HOMplot} for the fully indistinguishable photon inputs such that $P^{(U)}_n$ vanishes for $\varphi=\pi/n$ except for the special case of $n=3$.

$P^{(T)}_n$ never fully vanishes, instead it exhibits minima for $\varphi=\pi/2n$ that are given in table \ref{tableMinRelFr}. Thus full destructive interference is only observed for the case of fully indistinguishable input states with an even number of input photons.\\

\begin{figure}[!htb]
 \begin{center}
 \includegraphics[width=0.9\textwidth]{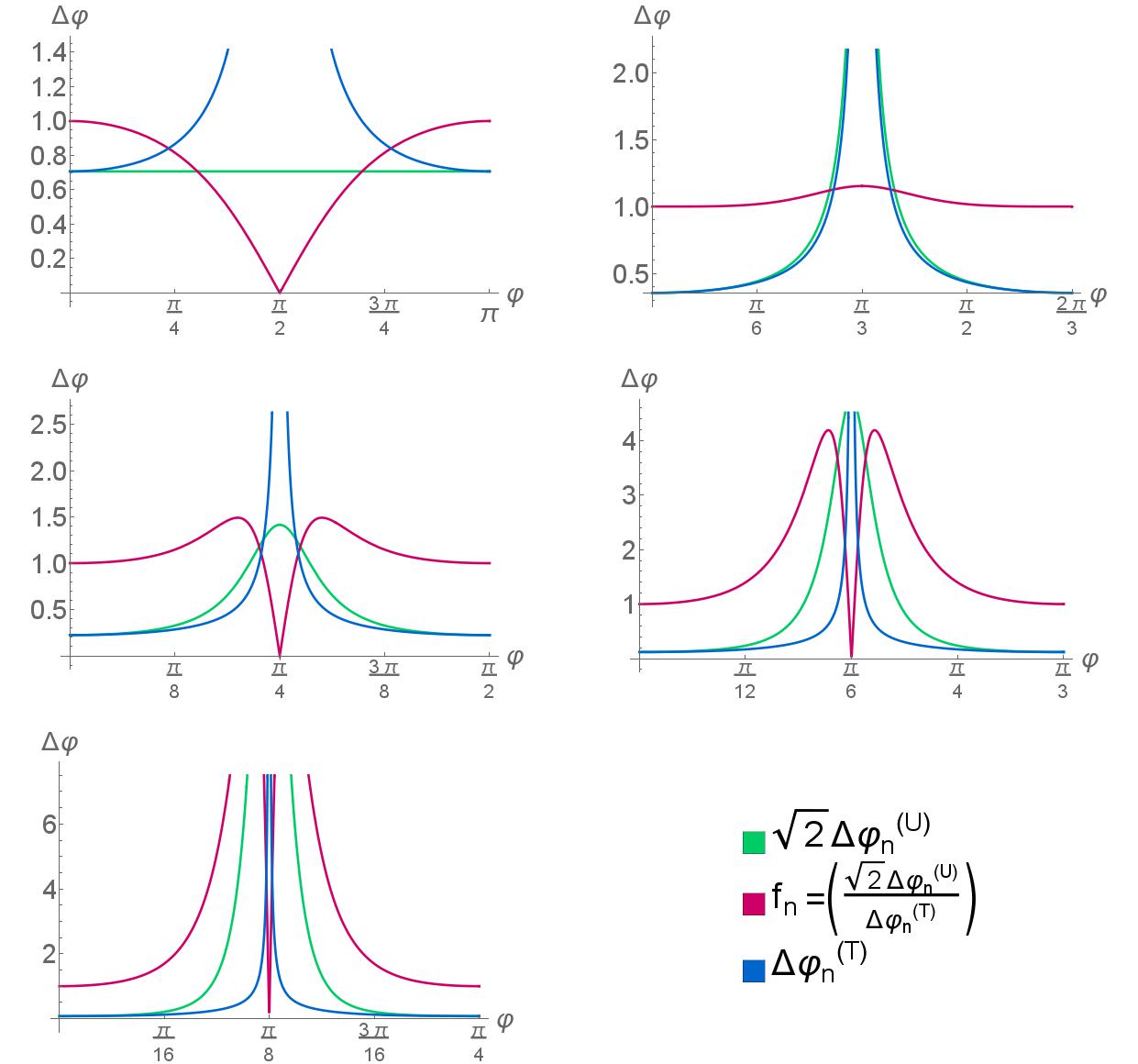}
\caption{ Comparison between the phase sensitivities $\Delta\varphi^{(T)}_n$ and $\Delta\varphi^{(U)}_n$ of the QuFTI given fully distinguishable (blue) and fully indistinguishable (green) input photons. The phase sensitivities are plotted over the phase $\varphi$. 
Additionally the ratio $f_n$ (red) of $\Delta\varphi^{(T)}_n$ and $\sqrt{2}\Delta\varphi^{(U)}_n$ is plotted. For better visibility the graphs have been restricted to their respective period. \label{figure:plotNOPhi} }
\end{center}
\end{figure}

Figure \ref{figure:plotNOPhi} compares the phase sensitivities $\Delta\varphi_n{(U)}$ and $\Delta\varphi_n{(T)}$ for indistinguishable and distinguishable photon inputs respectively. They are calculated from $P^{(U)}_n$ and $P^{(T)}_n$ according to eq. \eqref{Tphaseuncertainty}.
In the figure, $\Delta\varphi_n{(U)}$ is multiplied by a factor $\sqrt{2}$ to make the comparison of the phase sensitivities easier. 
The plot additionally shows the ratio $f_n={\sqrt{2}\Delta\varphi_n{(U)}}/{\Delta\varphi_n{(T)}}$ of the phase sensitivities.\\

The phase sensitivities are periodic functions with periods of  $2\pi/n$, depending on the respective values of the number $n$ of input photons. The plot clearly identifies the region of small phases around the periods as the regions of most precise measurements, given that the phase 
 sensitivities for both cases reach their minimal value. $f_n$ approaches the value 1 here for all values of $n$, showing that $\Delta\varphi_n{(U)}$ and $\Delta\varphi_n{(T)}$ only differ by a constant factor $\sqrt{2}$, that is independent of $n$. 
For different values of the phase $\varphi$ the ratio $f_n$ reaches values larger than one, indicating that the quantum advantage of the indistinguishable photon input is even smaller than $\sqrt{2}$. The exceptions are the dips of $f_n$ around $\varphi=2\pi/n\cdot k+\pi/n$ with $k\in\mathbb{N}$, where $\Delta\varphi_n{(T)}$ shows a pole and all phase information is consequently lost for measurements with distinguishable single photon inputs.
In the case of $n=3$ both $\Delta\varphi_n{(U)}$ and $\Delta\varphi_n{(T)}$ exhibit very similar behaviour, including the poles at $\varphi=\pi/3$. 
This indicates, that no true 3 photon interference is achieved in the QuFTI, but only interference of photon pairs.
In conclusion, the choice of indistinguishable photon inputs does not appear to produce an advantage in terms of precision against distinguishable photons, since the scaling is not influenced by a constant factor.

\section{Partial Distinguishability}\label{sec: docPO}

Another numerical approach to the QuFTI investigates the phase sensitivity $\Delta\varphi_n^{(PO))}$ (PO=partial overlap) of measurements given partially overlapping photonic spectra of the input photons. The probability $P_{n}^{(PO)}$ for this scenario is the averaged probability given by eq. \eqref{TMPIProbAverage} as
\begin{equation}
P_{n}^{(PO)}=\sum\limits_{\rho\in \Omega_N} f_\rho(S) \per A_\rho^{(D,S)}
\end{equation}
with the matrices 
$A_\rho^{(D,S)}=U^*_{d,s}U_{d,\rho(s)}$.\\

The script given in appendix \ref{APPENDIXMathematica} is employed to calculate the $P_{n}^{(PO)}$ for $n=2,3$ and $4$ shown in table \ref{tablePO}.
Each overlap factor $f_\rho$ defined by a permutation $\rho\in\Omega_n$ from the symmetric group $\Omega_n$ is subscripted by the set $\{1,2,...n\}$ after the application of the permutation.
From these $P_{n}^{(PO)}$ the corresponding phase sensitivities $\Delta\varphi_{n}^{(PO)}$ are derived according to eq. \eqref{Tphaseuncertainty} within the small phase approximations $\sin(\varphi)\approx \varphi$ and $\cos(\varphi)\approx 1-\varphi^2/2$, resulting in
\begin{align}
\Delta\varphi_{2}^{(PO)}=&\dfrac{1}{\sqrt{2}}\left(1+f_{\{2,1\}}\right)^{-1/2},
\label{DPhiPO2}
\end{align}
\begin{align}
\Delta\varphi_{3}^{(PO)}&=\frac{\sqrt{2}}{4}\left(1+ \frac{1}{3}(f_{\{1,3,2\}}+f_{\{2,1,3\}} +f_{\{3,2,1\}} )\right)^{-1/2}
\label{DPhiPO3}
\end{align}
and
\begin{align}
\Delta\varphi_{4}^{(PO)}&=\frac{1}{{2}}\left(5+  f_{\{1,2,4,3\}}+ f_{\{1,3,2,4\}}+f_{\{2,1,3,4\}}+f_{\{4,2,3,1\}}
 \right.\nonumber\\
&\left.\qquad+\frac{1}{2} \left( f_{\{1,4,3,2\}}+f_{\{3,2,1,4\}}
\right) \right)^{-1/2}.
\label{DPhiPO4}
\end{align}

\begin{table}
  \begin{center}
\begin{tabular}{|c||l|}
\hline
 \text{n} & $P_{n}^{(PO)}$ \\\hline\hline
 2 & $\frac{1}{4} (3+\text{Cos}[2 \varphi ]) f_{\{1,2\}}-\frac{1}{2} \text{Sin}[\varphi ]^2 f_{\{2,1\}}$ \\\hline
 3 & $\frac{1}{243} (147+92 \text{Cos}[3 \varphi ]+4 \text{Cos}[6 \varphi ]) f_{\{1,2,3\}} $\\&$
 +\frac{4}{243} (-6+5 \text{Cos}[3 \varphi ]+\text{Cos}[6
\varphi ]) \left( f_{\{1,3,2\}} + f_{\{2,1,3\}}+f_{\{3,2,1\}}\right)$\\&$
+\frac{32}{243} \text{Sin}\left[\frac{3
\varphi }{2}\right]^4 \left( f_{\{2,3,1\}}+ f_{\{3,1,2\}} \right)$\\\hline
 4 & $\frac{1805 \text{Cos}[4 \varphi ]+182 \text{Cos}[8 \varphi ]+3 (702+\text{Cos}[12 \varphi ])} {4096}f_{\{1,2,3,4\}}$\\&$
-\frac{(161+92 \text{Cos}[4
\varphi ]+3 \text{Cos}[8 \varphi ]) \text{Sin}[2 \varphi ]^2}{1024}
\left( f_{\{1,2,4,3\}}+ f_{\{1,3,2,4\}}+f_{\{2,1,3,4\}}+f_{\{4,2,3,1\}}
\right)$\\&$
-\frac{(81+44 \text{Cos}[4 \varphi ]+3 \text{Cos}[8 \varphi
]) \text{Sin}[2 \varphi ]^2 }{1024}\left( f_{\{1,4,3,2\}}+f_{\{3,2,1,4\}}
\right)$\\&$
+\frac{\text{Cos}[\varphi ]^4 (13+3 \text{Cos}[4 \varphi ]) \text{Sin}[\varphi ]^4}{16}  \left( f_{\{1,3,4,2\}}+f_{\{1,4,2,3\}} +f_{\{2,3,1,4\}}+f_{\{2,4,3,1\}}+f_{\{3,1,2,4\}}
\right)$\\&$
+\frac{(13+3 \text{Cos}[4 \varphi ]) \text{Sin}[2 \varphi ]^4}{256}
 \left(f_{\{2,1,4,3\}}+f_{\{3,2,4,1\}}
+f_{\{4,1,3,2\}}+ f_{\{4,2,1,3\}}
\right)$\\&$
+\frac{(-11+3 \text{Cos}[4 \varphi ]) \text{Sin}[2 \varphi ]^4}{256} \left( f_{\{2,3,4,1\}}
+f_{\{4,1,2,3\}}+f_{\{4,3,2,1\}}\right)$\\&$
+\frac{(1+3 \text{Cos}[4 \varphi
]) \text{Sin}[2 \varphi ]^4 }{256} \left( f_{\{2,4,1,3\}}+f_{\{3,1,4,2\}}+f_{\{3,4,1,2\}}+f_{\{3,4,2,1\}}
+f_{\{4,3,1,2\}}
\right)$\\
\hline
\end{tabular}
\end{center}
  \caption{Probabilities $P^{(PO)}_n $ for photons with partially overlapping photonic spectra distributions, computed by the \textit{Mathematica} code in Appendix \ref{APPENDIXMathematica}.\label{tablePO}}
\end{table}
 
Interestingly, permutations that interchange more than two indices in the set $\{1,2,...,n\}$ don't contribute to the result in the small phase approximation for the investigated $n$. 
The difference between no overlap and full overlap therefore comes only from the contributions of the permutations that only interchange two elements, that we call permutations of first order. 

The overlap factors are products of the normalized integrals over the spectral distributions of the input photons, as described by eq. \eqref{TMPIIndFac}.
Accordingly, they are all 0 for fully distinguishable input photons, in which case the only contributing permutation is the identity.
On the other hand, all $f_\rho$ are equal to 1 for fully indistinguishable photons and the permutations of first order contribute as well. 
Since the contribution of the identity is equal to the sum of the contributions of the first order permutations, this produces a factor $\sqrt{2}$ relating the cases of maximal and minimal overlap of the photonic spectra. 
This coincides with the results for full indistinguishability (full overlap) and full distinguishability (no overlap).\\



\chapter{Influence of Distinguishability in the Small Phase Approximation}\label{proof}

The present section introduces an analytical derivation of $P^{(U)}_n$ and $P^{(T)}_n$ in the small phase approximation for arbitrary $n$. In accordance with the results of the preceding chapter the resulting phase sensitivities $\Delta \varphi_n^{(U)}$ and $\Delta \varphi_n^{(T)}$ are shown to be related by a constant factor of $\sqrt{2}$.

\section{QuFTI Matrix Elements for Small Phases}\label{sec:FID}

To derive the QuFTI matrix $U^{(n)}$ for small phases, the approximation $e^\varphi\approx\varphi -\varphi/2$ is first applied to the phase shifter matrix $\Phi$ with the elements $\Phi_{j,k}$ given in eq. \eqref{TPhaseShifter}, neglecting contributions of cubic and higher order in $\varphi$.  
From eq.  \eqref{TMatrixElementsU} follow the matrix elements
\begin{align}
U_{j,k}^{(n)}&
\approx\sum\limits_{l=1}^n \frac{1}{n} e^{\frac{-2\im jl\pi}{n}}
\underbrace{\left(1+\im(l-1)\varphi- (l-1)^2\frac{\varphi^2}{2}\right)}_{\Phi_{j,k}} e^{\frac{2\im lk\pi}{n}}.
\label{proofElementsUSAA}
\end{align}

It is helpful to distinguish between the diagonal elements
\begin{align}
U_{j,j}^{(n)}&=
1+\im\varphi\frac{n-1}{2}- \frac{(n-1)(2n-1)}{6}\frac{\varphi^2}{2}
\label{proofUdiagonal}
\end{align}
and non-diagonal matrix elements 
\begin{align}
U_{j,k}^{(n)}&=
\left(\im\varphi+\varphi^2\right)\frac{1-e^{\frac{-2\im\pi(j-k)}{n}}}
{2-2\cos(\frac{2\pi(j-k)}{n})}
- \frac{1}{n}\sum\limits_{l=1}^n 
l^2 e^{\frac{-2\im l\pi}{n}(j-k)}\frac{\varphi^2}{2}
\label{proofUsides}
\end{align}
that are derived from eq. \eqref{proofUdiagonal} in the appendix \ref{appendixElementsUSAA}.

\section{Full Indistinguishability}
 
Next, the probability $P^{(U)}_n$ is calculated according to eq. \eqref{TprobFullOverlap}.
For this purpose it is useful to consider how different permutations contribute to the permanent, that is generally defined in \eqref{Tpermanent}. Using the diagonal elements in eq. \eqref{proofUdiagonal} the identity ($l=\sigma(l)\forall l$) contributes the value  
\begin{align}
\prod\limits_{l=1}^{n} U^{(n)}_{l,l}&=\left(1+\im\varphi\frac{n-1}{2}- \frac{(n-1)(2n-1)}{6}\frac{\varphi^2}{2} \right)^n
\nonumber\\
&\approx
1+n\frac{n-1}{2}\im\varphi- n\frac{(n-1)(2n-1)}{6}\frac{\varphi^2}{2}-\frac{n(n-1)}{2}\frac{(n-1)^2}{4}\varphi^2
\nonumber\\
&=1+n\frac{n-1}{2}\im\varphi- n\frac{(n-1)(3n^2-2n+1)}{12}\frac{\varphi^2}{2}.
\label{proofUcontrid}
\end{align}
to $\per U^{(n)}$.\\

A permutation $\sigma$ of first order in the symmetric group $\Omega_n$ only interchanges two indices $l_1$ and $l_2$ and therefore contributes the product of the two non diagonal matrix elements $U^{(n)}_{l_1,l_2}$ and $U^{(n)}_{l_2,l_1}$
with the remaining $n-2$ diagonal elements to the permanent. 
Neglecting terms above second order in $\varphi$ the product of the interchanged elements becomes
\begin{align}
U^{(n)}_{l_1,l_2}U^{(n)}_{l_2,l_1}=&-\varphi^2\frac{1-e^{\frac{-2\im\pi(l_1-l_2)}{n}}}
{2-2\cos(\frac{2\pi(l_1-l_2)}{n})}\cdot\frac{1-e^{\frac{-2\im\pi(l_2-l_1)}{n}}}
{2-2\cos(\frac{2\pi(l_2-l_1)}{n})}+O(\varphi^3)\nonumber\\
=&
-\varphi^2\frac{2-2\cos(\frac{2\pi(l_1-l_2)}{n})}
{(2-2\cos(\frac{2 \pi(l_1-l_2)}{n}))^2}=
-\frac{\varphi^2}{2}\frac{1}
{1-\cos(\frac{2\pi(l_1-l_2)}{n})}.
\label{proofU2elementsproduct}
\end{align}
Since $U^{(n)}_{l_1,l_2}U^{(n)}_{l_2,l_1}$ is already of second order it is sufficient to multiply it with the constant terms of the diagonal elements, that fortunately are all equal to 1.
Accordingly eq. \eqref{proofU2elementsproduct} is already the whole contribution of the permutation.\\

The symmetric group $\Omega_n$ contains $1/2\cdot n(n-1)$ permutations of first order for all possible pairs of indices from the set $\{1,2,...,n\}$. Their total contribution to $\per U^{(n)}$ is 
\begin{equation}
\frac{1}{2}\sum\limits_{l_1=1}^n\sum\limits_{\stackrel{l_2=1}{l_2\neq l_1}}^n U^{(n)}_{l_1,l_2}U^{(n)}_{l_2,l_1}
=
-\frac{\varphi^2}{4}\sum\limits_{l_1=1}^n\sum\limits_{\stackrel{l_2=1}{l_2\neq l_1}}^n \frac{1}{1-\cos(2\pi\frac{l_2-l_1}{n})}
=-\frac{\varphi^2}{4} \sum\limits_{l_1=1}^{n}\sum\limits_{\stackrel{l=1-l_1}{l\neq 0}}^{n-l_1} \frac{1}{1-\cos(2\pi\frac{l}{n})}
.
\label{proofU2permtotal1}
\end{equation}
Here the factor $1/2$ is multiplied because the contribution of each permutation is counted twice in the representation with two sums. With the symmetry of the cosine, the inner sum now contains all possible values for $l$ from 1 to $n-1$, independent of $l_2$. Therefore eq. \eqref{proofU2permtotal1} is equal to
\begin{align}
-&\frac{\varphi^2}{4}\sum\limits_{l_1=1}^n\sum\limits_{l_2=1}^{n-1} \frac{1}{1-\cos(2\pi\frac{l_2}{n})}=-\frac{n\varphi^2}{4}\sum\limits_{l_2=1}^{n-1} \frac{1}{1-\cos(2\pi\frac{l_2}{n})}
\nonumber\\
&=-\frac{n\varphi^2}{4}\frac{1}{2}\sum\limits_{l_2=1}^{n-1} \frac{1}{\sin^2(\pi\frac{l_2}{n})}\overset{\text{\eqref{TSinusId}}}{=}-\frac{n\varphi^2}{4}\frac{n^2-1}{6}
\label{proofU2permtotal2}
\end{align}
using the identity \eqref{TSinusId}. \\

The permutations of first order contribute terms at least of second order in $\varphi$ to the permanent. The contribution of  higher order permutations with $m$ interchanged elements is a product of $m$ terms in each of which the lowest order in $\varphi$ is linear. The lowest order in $\varphi$ of such a permutation is therefore
\begin{equation}
\frac{(\im\varphi)^m}{2^m}\prod\limits_{i=1}^m \frac{n\left(1-e^{\frac{-2\im \pi(l_i-\sigma(l_i))}{n}}\right)}
{1-\cos(\frac{2 \pi(l_i-\sigma(l_i))}{n})},
\label{proofmpower}
\end{equation}
which can safely be neglected in the small phase approximation if $m>2$. In second order in $\varphi$ therefore only the permutations of first order and the identity contribute at all to the permanent, that is
\begin{align}
\per U^{(n)}&=1+n\frac{n-1}{2}\im\varphi- n\frac{(n-1)(3n^2-2n+1)}{12}\frac{\varphi^2}{2}-\frac{n\varphi^2}{4}\frac{(n-1)(n+1)}{6}\nonumber\\
&=1+n\frac{n-1}{2}\im\varphi- n\frac{(n-1)(3n^2-n+2)}{12}\frac{\varphi^2}{2}.
\end{align}

The resulting probability $P^{(U)}_n$ for an $n$-fold detection given fully indistinguishable input photons is then
\begin{align}
&P^{(U)}_n=|\per U^{(n)}|^2
=1-\frac{n(n-1)(n+1)}{6}\varphi^2.
\label{proofFOResult}
\end{align}
This is the probability that was observed numerically in \cite{Motes} as given by eq. \eqref{TProbSAA}. \\

\section{Full Distinguishability}

For an input of fully distinguishable photons, the probability of detecting one photon per output mode $P^{(T)}_n$ 
is given by the permanent of the matrix $T^{(n)}$ with the matrix elements defined by eq. \eqref{TprobNoOverlap}. The diagonal elements
\begin{align}
T^{(n)}_{j,j}&=|U^{(n)}_{j,j}|^2=\left|1+\im\varphi\frac{n-1}{2}- \frac{(n-1)(2n-1)}{6}\frac{\varphi^2}{2}\right|^2
\nonumber\\
&\approx1+\varphi^2\frac{(n-1)^2}{4}- \frac{(n-1)(2n-1)}{6}\varphi^2+O(\varphi^3)
\nonumber\\
&=
1-\frac{(n-1)(n+1)}{12}\varphi^2
\end{align}
follow from eq. \eqref{proofUdiagonal} and the non diagonal elements 
\begin{align}
T^{(n)}_{j,k}&=|U^{(n)}_{j,k}|^2=-\varphi^2\frac{1-e^{\frac{-2\im l\pi(j-k)}{n}}}
{2-2\cos(\frac{2 l\pi(j-k)}{n})}\cdot\frac{1-e^{\frac{2\im l\pi(j-k)}{n}}}
{2-2\cos(\frac{2 l\pi(j-k)}{n})}
\nonumber\\
&=-\frac{\varphi^2}{2}\frac{1}
{1-\cos(\frac{2 l\pi(j-k)}{n})}.
\end{align}
follow from eq. \eqref{proofUsides}.\\

The product of two non-diagonal elements is
\begin{align}
T^{(n)}_{j_1,k_1}T^{(n)}_{j_2,k_2}=\dfrac{\varphi^4}{4\cdot(1-\cos({2\pi\frac{j_1-k_1}{n}}))\cdot(1-\cos({2\pi\frac{j_2-k_2}{n}}))},
\end{align}
which is a term of fourth order in $\varphi$.
It can therefore be safely deduced, that in second order in $\varphi$ no permutation but the identity contributes at all to the permanent. The probability $P^{(U)}_n$ in the small phase approximation is therefore simply
\begin{align}
P^{(T)}_n=\per T^{(n)}\approx\prod\limits_{l=1}^n T_{l,l}^{(n)}= (T_{1,1}^{(n)})^n=1-\frac{n(n-1)(n+1)}{12}\varphi^2.
\label{proofNOResult}
\end{align}

\section{Partial Distinguishability}

In the most general case of input photons with partially overlapping photonic spectra, the probability $P_{n}^{(PO)}$ for an $n$-fold detection is given by eq. \eqref{TMPIProbAverage}.
For a given permutation $\rho\in\Omega_n$ the matrix elements of the corresponding matrix  $A_\rho^{(D,S)}=A^{(n)}_\rho$ follow from the elements of $U^{(n)}$ in eq. \eqref{proofUdiagonal} and eq. \eqref{proofUsides}. 
The diagonal elements of $A_\rho$ are 
\begin{equation}
A^{(n)}_{\rho;j,j} = \begin{cases} T_{j,j}=1-\frac{(n-1)(n+1)}{12}\varphi^2 &\mbox{if } \rho(j)= j \\
U^*_{j,j}U_{j,\rho(j)}= \left(\im\varphi+\frac{n+1}{2}\varphi^2\right)\frac{1-e^{\frac{-2\im\pi(j-\rho(j))}{n}}}
{2-2\cos(\frac{2\pi(j-\rho(j))}{n})}- \frac{\varphi^2}{2n}\sum\limits_{l=1}^n 
l^2 e^{\frac{-2\im l\pi(j-\rho(j))}{n}}
& \mbox{if } \rho(j)\neq j \end{cases}
\label{proofAdiag}
\end{equation}
and the non diagonal elements are
\begin{equation}
A^{(n)}_{\rho;j,k} = \begin{cases} T_{j,k}=-\frac{\varphi^2}{4}\frac{1}
{1-\cos(\frac{2 l\pi(j-k)}{n})} &\mbox{if } \rho(k)= k \\
U^*_{j,k}U_{j,j}= \left(-\im\varphi+\frac{n+1}{2}\varphi^2\right)\frac{1-e^{\frac{2\im\pi(j-k)}{n}}}
{2-2\cos(\frac{2\pi(j-k)}{n})}-\frac{\varphi^2}{2n}\sum\limits_{l=1}^n 
l^2 e^{\frac{-2\im l\pi(j-k)}{n}}
& \mbox{if } \rho(k)= j \\
U^*_{j,k}U_{j,\rho(k)}= 
\varphi^2\frac{1-e^{\frac{2\im\pi(j-k)}{n}}}
{2-2\cos(\frac{2\pi(j-k)}{n})}
\cdot\frac{1-e^{\frac{-2\im\pi(j-\rho(k))}{n}}}
{2-2\cos(\frac{2\pi(j-\rho(k))}{n})}
& \mbox{if } \rho(k)\neq j,k.
 \end{cases}
 \label{proofAnondiag}
\end{equation}

 Considering that only diagonal elements that fulfil $\rho(l)=l$ contain constant terms, the permanent of a matrix $A^{(n)}_\rho$ for a given $\rho$ that interchanges $m$ indices would be at least of the order $m$. Therefore only $A^{(n)}_\rho$ defined by permutations $\rho$ of first order or the identity contribute to the probability $P_{n}^{(PO)}$ if terms of order $\varphi^3$ and higher are neglected. The contribution of the identity is simply the permanent of $T^{(n)}$ in eq. \eqref{proofNOResult}. \\

If $\rho$ is a permutation of first order, it is again useful to consider the contributions of different permutations $\sigma$ in calculating $\per A_\rho^{(n)}$.
Firstly, the contribution of the identity is given by the product of all diagonal elements. 
The product of the two elements interchanged by $\rho$ is 
\begin{align}
A^{(n)}_{\rho;l_1,l_1}A^{(n)}_{\rho;l_2,l_2}=&-\varphi^2
\frac{1-e^{\frac{-2\im\pi(l_1-\rho(l_1))}{n}}}
{2-2\cos(\frac{2\pi(l_1-\rho(l_1))}{n})}\frac{1-e^{\frac{-2\im\pi(l_2-\rho(l_2))}{n}}}
{2-2\cos(\frac{2\pi(l_2-\rho(l_2))}{n})}\nonumber\\
=&-\varphi^2
\frac{1-e^{\frac{-2\im\pi(l_1-l_2)}{n}}}
{2-2\cos(\frac{2\pi(l_1-l_2)}{n})}\frac{1-e^{\frac{-2\im\pi(l_2-l_1)}{n}}}
{2-2\cos(\frac{2\pi(l_2-l_1)}{n})}
\nonumber\\
=&-\varphi^2
\frac{1}
{2-2\cos(\frac{2\pi(l_1-l_2)}{n})}
\label{proofA2elementsproduct}
\end{align}
which is equal to \eqref{proofU2elementsproduct}. 
As this term is already of second order in $\varphi$, it is sufficient to multiply it with the constant terms of the remaining diagonal elements. 
Since these are all equal to 1 the product of the two elements eq. \eqref{proofA2elementsproduct} is already the total contribution of the identity. \\

The contribution of an arbitrary permutation $\sigma$ of first order similarly follows from the product of the two interchanged non-diagonal elements
\begin{align}
A^{(n)}_{\rho;l_1,\sigma(l_1)}A^{(n)}_{\rho;l_2,\sigma(l_2)}&=U^*_{l_1,\sigma(l_1)}U_{l_1,\rho(\sigma(l_1))}\cdot U^*_{l_2,\sigma(l_2)}U_{l_2,\rho(\sigma(l_2))}\nonumber\\
&=U^*_{l_1,l_2}U_{l_1,\rho(l_2)}\cdot U^*_{l_2,l_1}U_{l_2,\rho(l_1)}.
\label{proofAdiagprod0}
\end{align}
According to eq. \eqref{proofAnondiag} the product yields terms of order $\varphi^3$ and higher except if $\rho(l_2)=l_1$ and $\rho(l_1)=l_2$.
 In this case $\sigma$ and $\rho$ describe the same permutation and eq. \eqref{proofAdiagprod0} becomes equal to eq. \eqref{proofA2elementsproduct} and the same term is contributed twice to the permanent.
Thus, $\per A^{(n)}_\rho$ is
\begin{equation}
\per A^{(n)}_\rho\approx 
-\varphi^2\frac{1}
{1-\cos(\frac{2\pi(l_1-l_2)}{n})},
\end{equation}
if $\rho$ is a permutation of first order.\\

Consequently, $P_{n}^{(PO)} $ is
\begin{align}
P_{n}^{(PO)}&=1-\frac{n(n-1)(n+1)}{12}\varphi^2-\frac{\varphi^2}{2} \sum\limits_{l_1=1}^n\sum\limits_{\stackrel{l_2=1}{l_2\neq l_1}}^n \frac{f_{\{l_1,l_2\}}}
{1-\cos(\frac{2\pi(l_1-l_2)}{n})}
\nonumber\\&=1-\frac{n(n-1)(n+1)}{12}\varphi^2-\frac{\varphi^2}{4} \sum\limits_{l_1=1}^n\sum\limits_{\stackrel{l_2=1}{l_2\neq l_1}}^n \frac{f_{\{l_1,l_2\}}}
{\sin^2(\frac{\pi(l_1-l_2)}{n})}
\label{POresult}
\end{align}
where each permutation $\rho=\{l_1,l_2\}$ of first order is denoted by the two indices it interchanges. The sum contains all $n(n-1)/2$ permutations of first order twice and is therefore multiplied by a factor $1/2$.\\

If all overlap factors $f_\rho$ are zero for an input of fully distinguishable single photons the sum vanishes and the result matches eq. \ref{proofNOResult}. If on the other hand all $f_\rho$ are 1 which corresponds to the fully distinguishable case, the sum yields the result in eq. \eqref{proofFOResult}. 
Thus, the results for the partially distinguishable photon input match those of the two extreme cases of overlap.

\section{Resulting Phase Sensitivities}

Finally the phase sensitivities are derived from the respective probabilities of all examined cases of photon overlap. 
Equations \eqref{proofFOResult} and \eqref{proofNOResult} relate the probabilities for fully distinguishable and a fully indistinguishable photon input via
\begin{equation}
P^{(U)}_n=1- k(n)\varphi^2
\end{equation}
and
\begin{equation}
P^{(T)}_n=1-\frac{1}{2} k(n)\varphi^2.
\label{proofNOResultk(n)}
\end{equation}
 with $k(n)={n(n-1)(n+1)}/{6}$. 
The phase sensitivity $\Delta \varphi^{(U)}_n$ that is derived from the full photon overlap is recalled from eq. \eqref{TphaseSensitivitySAA}. Using eq. \eqref{proofNOResultk(n)} $\Delta \varphi^{(T)}_n$ is
\begin{align}
\Delta \varphi^{(T)}_n=\frac{\sqrt{P^{(T)}-(P^{(T)})^2}}{\left|\frac{\partial P^{(T)}}{\partial\varphi}\right|}
=\frac{\sqrt{1-\frac{1}{2}k(n)\varphi^2-1+k(n)\varphi^2}}{k(n)\left|\varphi\right|}
=\frac{1}{\sqrt{2} \sqrt{ k(n)}}.
\label{ResultPhaseSens}
\end{align}
 The phase sensitivities for both cases are therefore related as
\begin{equation}
\Delta\varphi^{(T)}_n=\sqrt{2} \Delta\varphi^{(U)}_n
\label{proofresult}
\end{equation}
for small phases $\varphi$. 
With eq. \eqref{POresult} the phase sensitivity $\Delta \varphi^{(PO)}_n$ for the partially distinguishable photons is
\begin{equation}
\Delta \varphi^{(PO)}_n=\frac{\sqrt{P^{(PO)}-(P^{(PO)})^2}}{\left|\frac{\partial P^{(PO)}}{\partial\varphi}\right|}
=\frac{1}{\sqrt{2}} \left(k(n)-\frac{1}{2} \sum\limits_{l_1=1}^n\sum\limits_{\stackrel{l_2=1}{l_2\neq l_1}}^n \frac{f_{\{l_1,l_2\}}}
{\sin^2(\frac{\pi(l_1-l_2)}{n})}
\right)^{-\frac{1}{2}}.
\end{equation}

In conclusion the analytical investigation on the influence of distinguishability on the QuFTI provides an analytical derivation of the pattern for $P^{(U)}_n$ observed in \cite{Motes}. Most importantly, the interferometric phase sensitivities for fully indistinguishable and fully distinguishable input photons are proven to be equal apart from a constant factor $\sqrt{2}$, independent of the number $n$ of input photons. 
Surprisingly, this constant factor represents the only enhancement in phase sensitivity arising from the occurrence of multi photon interference and entanglement generation in the QuFTI.

\chapter{Conclusions}

This work investigates how the quantum Fourier transform interferometer (QuFTI) is influenced by different degrees of indistinguishability of the $n$ input photons. The results of the numerical approach in section \ref{sec: doc} indicate, that a lower degree of distinguishability does not necessarily correspond to a lower precision of the measurement. It is recalled from figure \ref{figure:plotNOPhi}, which compares the phase sensitivity for fully distinguishable and fully indistinguishable photons, that the opposite can even be the case for certain values of the phase.
Furthermore, for small phase values the interferometric phase sensitivities are demonstrated analytically to be equal apart from a constant factor $\sqrt{2}$ for an arbitrary number $n$ of input photons. \\

In conclusion, this constant factor $\sqrt{2}$ represents the only enhancement in phase sensitivity that can be obtained by exploiting multi-photon quantum interference and entanglement generation in the QuFTI with respect to the "classical" case of photons completely distinguishable at the detectors.
A consequence of this important result is, that the resource counting procedure in \cite{Motes} can not be correct and shot-noise limit can only be beaten for a small number of single photons. 

\let\cleardoublepage\clearpage

\appendix
\chapter{Trigonometric Series}

Firstly consider the spread polynomials $S_m(x)$ that are defined using the Chebyshev polynomials \cite[p. 951f]{Bronstein} $T_m(x)$ as
\begin{equation}
S_m(x)=\frac{1-T_m(1-2x)}{2}.
\end{equation}

The $T_m(x)$ fulfil the property $T_m(\cos(x))=\cos(m x)$, therefore the polynomials $S_m$ must satisfy
\begin{align}
S_m(\sin^2(x))&=\frac{1-T_m(1-2\sin^2(x))}{2}=\frac{1-T_m(\cos(2x))}{2}\nonumber\\
&=\frac{1-\cos(2mx)}{2}=\frac{2\sin^2(mx)}{2}=\sin^2(mx).
\end{align}

Therefore it follows
\begin{align}
S_{m+1}&(\sin^2(x))+S_{m-1}(\sin^2(x))=\sin^2(mx+x)+\sin^2(mx-x)\nonumber\\
&=(\sin(mx)\cos(x)+\cos(mx)\sin(x))^2\nonumber\\
&\qquad
+(\sin(mx)\cos(x)-\cos(mx)\sin(x))^2\nonumber\\
&=2\sin^2(mx)\cos^2(x)+2\cos^2(mx)\sin^2(x) \nonumber\\
&=2(1-\sin^2(x))S_m(\sin^2(x))+2\sin(x)(1-S_m(\sin^2(x)))
\end{align}
which yields

\begin{equation}
S_{m+1}(x)=2(1-2x)S_{m}(x)-S_{m-1}(x)+2x.
\label{TidSm+1}
\end{equation}

Now because of the relation
\begin{equation}
S_m(\sin^2(\frac{k\pi}{m}))=\sin^2(m\frac{k\pi}{m})=0 \mathrm{\qquad for\qquad} k\in \{0,1,...,n-1\}
\end{equation}

the $\sin^2(\frac{k\pi}{m})$ with $k\in \{0,1,2,...,n-1\}$ are roots of the polynomial $S_m$. Further set $S_m(x)=x P_m(x)$
with $ P_m(x)= a_m+xb_m+x^2 Q(x)$
such that $\sin^2(\frac{k\pi}{m})$ with $k\in \{1,2,...,n-1\}$ are the polynomial roots of $P_m(x)$ and eq. \eqref{TidSm+1} becomes

\begin{equation}
P_{m+1}(x)=2(1-2x)P_{m}(x)-P_{m-1}(x)+2.
\label{TidPm+1}
\end{equation}

Now by calculating the common denominator and using Vieta's Formula \cite[p. 44]{Bronstein} the sum can be transferred via
\begin{equation}
\sum\limits_{l=1}^{n-1}\frac{1}{\sin^2(\frac{\pi l}{n})}=\frac{\sum\limits_{l=1}^{n-1} \prod\limits_{k\neq l} \sin^2(\frac{\pi l}{n})}{\prod\limits_{k=1}^{n-1} \sin^2(\frac{\pi l}{n})}
=\frac{s_{n-2}}{s_{n-1}}=-\frac{b_m}{a_m}.
\label{Tidsum}
\end{equation}

Now it is finally shown by induction that 
\begin{equation}
a_m=m^2 \mathrm{\qquad and\qquad}b_m=-\frac{m^2(m^2-1)}{3}.
\end{equation}

\paragraph{1.} For $m=0$ it follows with the property $T_0(x)=1$ \cite[p. 951f]{Bronstein}
\begin{equation}
P_0(x)=\frac{1}{x} S_0(x)=\frac{1}{x}\frac{1-T_0(1-2x)}{2}=0.
\end{equation} 

\paragraph{2.} With $T_1(x)=x$ the case $m=1$ yields
\begin{equation}
P_1(x)=\frac{1}{x} S_1(x)=x\frac{1-1+2x}{2}=1.
\end{equation}

\paragraph{3.} Now assume that the proposition is true for one $m$ and $m-1$, then it follows according to eq. \eqref{TidPm+1} that

\begin{align}
P_{m+1}(x)&=2(1-2x)P_{m}(x)-P_{m-1}(x)+2\nonumber\\
&= 2(1-2x)\left( a_m+b_m x+x^2Q_m(x)  \right)
\nonumber\\&\qquad
 -\left( a_{m-1}+b_{m-1}x+x^2Q_{m-1}(x)  \right)+2
 \nonumber\\
 &=(2+2a_m-a_{m-1})+x(2b_m-4a_m-b_{m-1})
 \nonumber\\&\qquad
 +x^2(4b_m+Q(x)_m-Q(x)_{m-1}).
\end{align}

From which it can be identified that
\begin{align}
a_{m+1}&=2+2a_m-a_{m-1}=2+2m^2-(m-1)^2\nonumber\\
&=2+2m^2-m^2+2m-1=m^2+2m+1=(m+1)^2
\end{align}
and
\begin{align}
b_{m+1}&=2b_m-4a_m-b_{m-1}
=-2\frac{m^2(m^2-1)}{3}-4m^2+\frac{(m-1)^2((m-1)^2-1)}{3}
\nonumber\\
&=-\frac{2m^4-2m^2+12m^2-(m^2-2m+1)(m^2-2m)}{3}
\nonumber\\
&=-\frac{m^4+4m^3+5m^2+2m}{3}
=-\frac{(m+1)^2((m+1)^2-1)}{3}.
\end{align}

Such that finally the sum \eqref{Tidsum} can be written as

\begin{equation}
\sum\limits_{l=1}^{n-1}\frac{1}{\sin^2(\frac{\pi l}{n})}=\frac{m^2(m^2-1)}{3m^2}=\frac{m^2-1}{3}.
\end{equation}

\chapter{Calculations}

%
%


\section{$P^{(U)}_n$ in the small angle approximation}\label{appendixPUn}

The probability $P^{(U)}_n$ becomes in the small angle approximation

\begin{align}
 P^{(U)}_n=&\frac{1}{n^{2n-2}} \prod\limits_{j=1}^{n-1} \left[a_n(j)\cos(n\varphi)+b_n(j)\right]
 \approx
 \frac{1}{n^{2n-2}} \prod\limits_{j=1}^{n-1} \left[ a_n(j)(1-\frac{n^2}{2}\varphi^2)+b_n(j) \right]
\nonumber \\=&
 \frac{1}{n^{2n-2}} \prod\limits_{j=1}^{n-1} \left[ n^2-n^2\frac{a_n(j)}{2}\varphi^2 \right]
=
\prod\limits_{j=1}^{n-1}\left[ 1-(nj-j^2)\varphi^2\right]
\nonumber\\=&
 1-\varphi^2\sum\limits_{j=1}^{n-1} \left[ nj-j^2\right] +O(\varphi^4)
\overset{\text{\eqref{TgaussianSum}}}{=} 1-\varphi^2\frac{n^2(n-1)}{2}+\varphi^2\sum\limits_{j=1}^{n-1} j^2\nonumber\\
\overset{\text{\eqref{TgaussianSum(2)}}}{=}&
1-\varphi^2\left(n\frac{n(n-1)}{2}-\frac{n(n-1)(2n-1)}{6}\right)=
1-\varphi^2\frac{n(n-1)(n+1)}{6}
\end{align}
using the identities \eqref{TgaussianSum} and \eqref{TgaussianSum(2)}.

\section{Elements of $U^{(n)}$ in the small angle approximation}\label{appendixElementsUSAA}
The diagonal elements of the matrix $U{(n)}$ in the small angle approximation follow from eq. \eqref{proofElementsUSAA} as

\begin{align}
U_{j,j}^{(n)}&=
\sum\limits_{l=1}^n \frac{1}{n} e^{\frac{-2\im jl\pi}{n}}\left(1+\im(l-1)\varphi- (l-1)^2\frac{\varphi^2}{2}\right) e^{\frac{2\im lk\pi}{n}} 
\nonumber\\
&=
\sum\limits_{l=1}^n \frac{1}{n} \left(1+\im(l-1)\varphi- (l-1)^2\frac{\varphi^2}{2}\right)
\nonumber\\
&=
\sum\limits_{l=1}^n \frac{1}{n} \left(1-\im\varphi-\frac{\varphi^2}{2}\right)
+
\sum\limits_{l=1}^n \frac{l}{n} \left(\im\varphi+\varphi^2\right)
-
\sum\limits_{l=1}^n \frac{l^2}{n}\frac{\varphi^2}{2}
\nonumber\\
&\overset{\text{\eqref{TgaussianSum}}}{=}
1-\im\varphi-\frac{\varphi^2}{2}
+
\frac{n+1}{2} \left(\im\varphi+\varphi^2\right)
-
\sum\limits_{l=1}^n \frac{l^2}{n}\frac{\varphi^2}{2}
\nonumber\\
&\overset{\text{\eqref{TgaussianSum(2)}}}{=}
1-\im\varphi-\frac{\varphi^2}{2}+
\frac{n+1}{2} \left(\im\varphi+\varphi^2\right)- \frac{(n+1)(2n+1)}{6}\frac{\varphi^2}{2}
\nonumber\\
&=
1+\im\varphi\frac{n-1}{2}- \frac{(n-1)(2n-1)}{6}\frac{\varphi^2}{2}
\label{APPENDIXproofUdiagonal}
\end{align}
using the identities \eqref{TgaussianSum} and \eqref{TgaussianSum(2)}. The non-diagonal elements follow again from eq. \eqref{proofElementsUSAA} as

\begin{align}
U_{j,k}^{(n)}&=
\sum\limits_{l=1}^n V_{j,l}\Phi_{l,l}V_{l,k}^\dagger\frac{1}{n} e^{\frac{-2\im jl\pi}{n}}\left(1+\im(l-1)\varphi- (l-1)^2\frac{\varphi^2}{2}\right) e^{\frac{2\im lk\pi}{n}} 
\nonumber\\
&=\frac{1}{n}\sum\limits_{l=1}^n 
e^{\frac{-2\im l\pi}{n}(j-k)}\left(1+\im(l-1)\varphi- (l-1)^2\frac{\varphi^2}{2}\right) 
\nonumber\\
&=\frac{1}{n}\sum\limits_{l=1}^n 
e^{\frac{-2\im l\pi}{n}(j-k)}\left(1-\im\varphi-\frac{\varphi^2}{2}\right)
 + \frac{1}{n}\sum\limits_{l=1}^n 
l e^{\frac{-2\im l\pi}{n}(j-k)}\left(\im\varphi+\varphi^2\right)
 \nonumber\\
 &\qquad- \frac{1}{n}\sum\limits_{l=1}^n 
l^2 e^{\frac{-2\im l\pi}{n}(j-k)}\frac{\varphi^2}{2}.
\label{APPENDIXproofUnondiag}
\end{align}

Here the first sum can be calculated with the geometric series \eqref{TgeoSeries}. It then yields
\begin{equation}
\sum\limits_{l=1}^n e^{\frac{-2\im l\pi}{n}(j-k)}
=\sum\limits_{l=0}^{n-1} \left(e^{\frac{-2\im\pi}{n}(j-k)}\right)^l
\overset{\text{\eqref{TgeoSeries}}}{=}\frac{1-\left(e^{\frac{-2\im\pi}{n}(j-k)}\right)^n}{1-e^{\frac{-2\im\pi}{n}(j-k)}}
=\frac{1-e^{-2\im\pi(j-k)}}{1-e^{\frac{-2\im\pi}{n}(j-k)}}.
\end{equation}

The index shift was possible, because the $n$\textsuperscript{th} element is equal to the 0\textsuperscript{th}. Furthermore, as $j-k$ must be a whole number, the term $e^{\frac{-2\im\pi}{n}(j-k)}$ on the far right side will become equal to 1 and therefore the whole sum vanishes. The second sum can be simplified as well using the identity \eqref{TgeoSeries(2)}. It then yields
\begin{align}
\sum\limits_{l=1}^n& l e^{\frac{-2\im l\pi}{n}(j-k)}
=\sum\limits_{l=0}^n l \left(e^{\frac{-2\im \pi}{n}(j-k)}\right)^l
\nonumber\\
\overset{\text{\eqref{TgeoSeries(2)}}}{=}&\frac{ne^{\frac{-2\im \pi(j-k)}{n} (n+2)} -(n+1) e^{\frac{-2\im \pi(j-k)}{n}(n+1)}+e^{\frac{-2\im \pi(j-k)}{n}} }
{\left(1-e^{\frac{-2\im \pi(j-k)}{n}}\right)^2}\nonumber\\
&=\frac{-ne^{\frac{-2\im \pi(j-k)}{n}}}
{1-e^{\frac{-2\im \pi(j-k)}{n}}}
=\frac{-ne^{\frac{-2\im \pi(j-k)}{n}}\left(1-e^{\frac{2\im \pi(j-k)}{n}}\right)}
{2-2\cos(\frac{2 \pi(j-k)}{n})}
=\frac{n\left(1-e^{\frac{-2\im \pi(j-k)}{n}}\right)}
{2-2\cos(\frac{2 \pi(j-k)}{n})}.
\end{align}

Equation \eqref{APPENDIXproofUnondiag} becomes accordingly

\begin{align}
U_{j,k}^{(n)}&=
\left(\im\varphi+\varphi^2\right)\frac{1-e^{\frac{-2\im\pi(j-k)}{n}}}
{2-2\cos(\frac{2\pi(j-k)}{n})}
- \frac{1}{n}\sum\limits_{l=1}^n 
l^2 e^{\frac{-2\im l\pi}{n}(j-k)}\frac{\varphi^2}{2}.
\end{align}

\chapter{\textit{Mathematica} Codes}\label{APPENDIXMathematica}

\includepdf[pages=-]{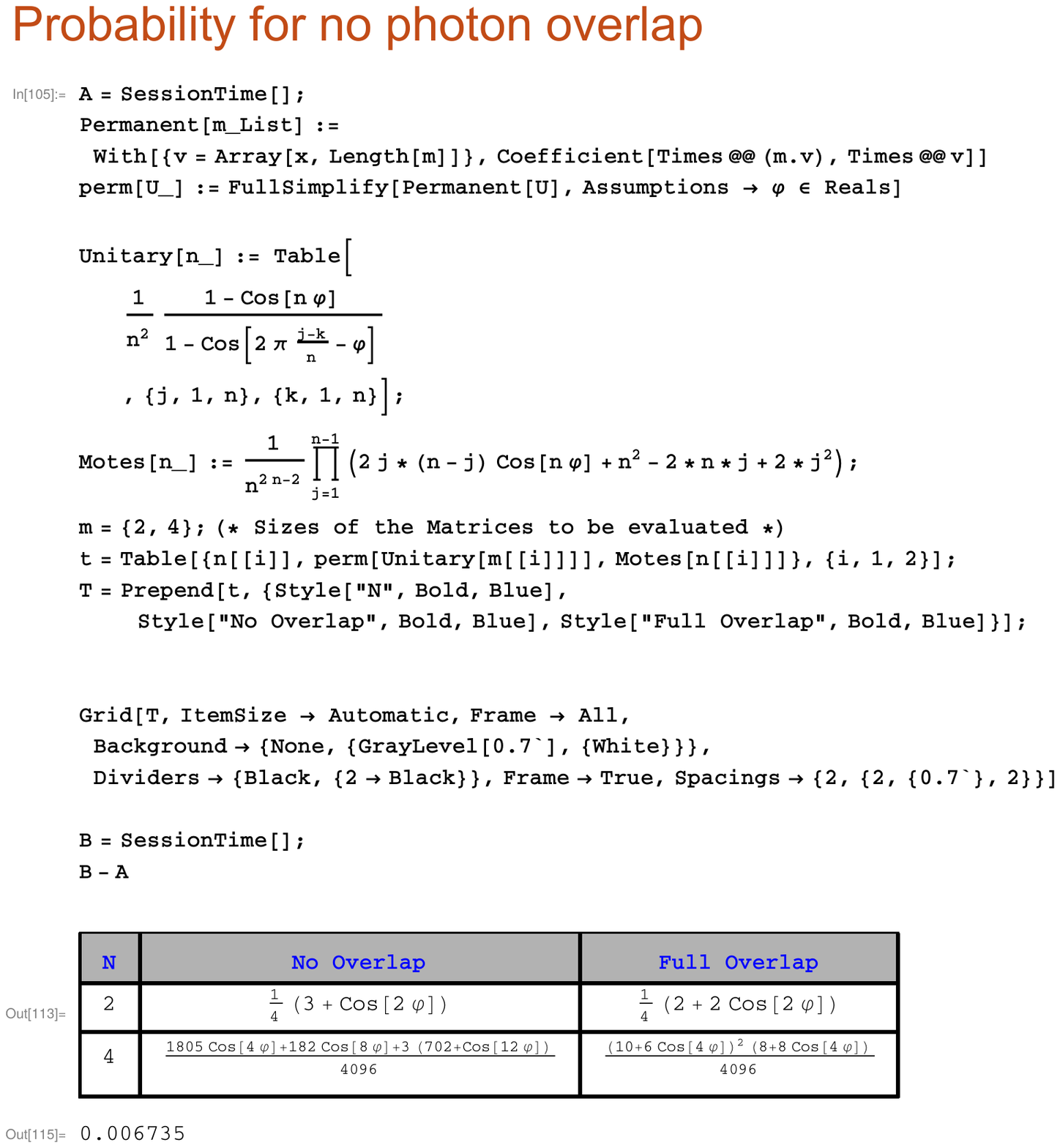}
\includepdf[pages=-]{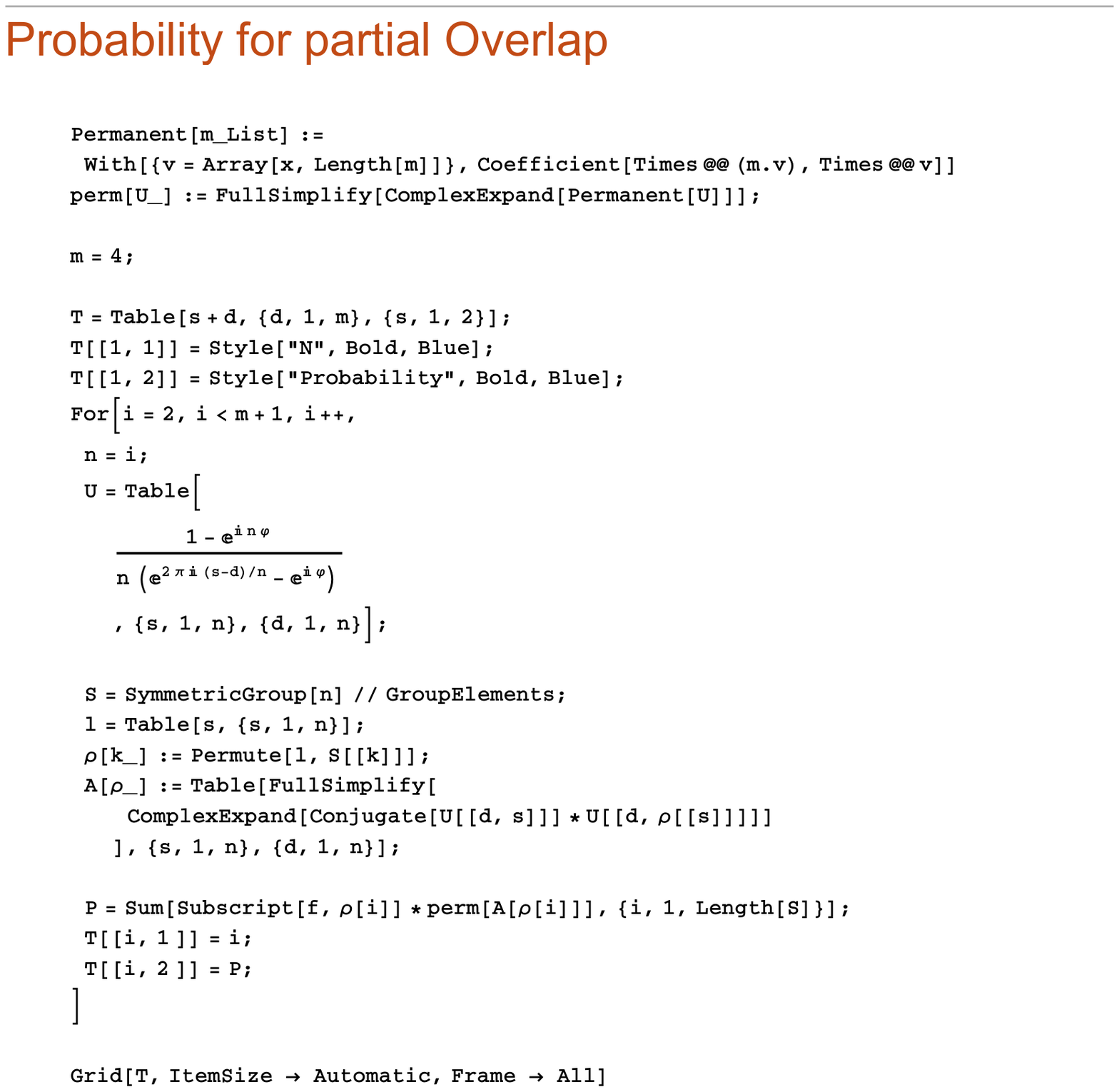}

\begingroup
\listoffigures
\addcontentsline{toc}{chapter}{List of Figures}
\let\clearpage\relax
\let\cleardoublepage\relax
\listoftables
\addcontentsline{toc}{chapter}{List of Tables}
\endgroup

\pagestyle{plain}

\section*{Acknowledgements}
\addcontentsline{toc}{chapter}{Acknowledgements}
I would like to thank Prof. Dr. Wolfgang Schleich and Dr. Vincenzo Tamma for giving me the opportunity to work in a very up-to-date topic and to explore the world of quantum optics and metrology. Furthermore I would like to thank the whole staff of the department of quantum physics for a very welcoming atmosphere. Very special thanks goes to Simon Laibacher for his advice and help with my thesis.\\

I also would like to thank those people who hunted down my mistakes while proofreading my thesis: Janica, Fabian, Annika, Stefanie, Rachel, Nat and Dipl.-Jur. S{\"o}ren Zimmermann.\\

I dedicate this thesis to my parents for all their love and support.\\

\end{document}